\documentclass[twocolumn]{aastex63}

\usepackage{bm}
\received{...}
\revised{...}
\accepted{...}
\submitjournal{ApJ}

\shorttitle{Gaussian Process Modeling $\gamma$-ray Blazar Variability}
\shortauthors{Yang et al.}

\begin{document}
	
	\title{Gaussian Process Modeling \textit{Fermi}-LAT $\gamma$-ray Blazar Variability:  A Sample of Blazars with $\gamma$-ray Quasi-periodicities}

	\correspondingauthor{Dahai Yan}
	\email{yandahai@ynao.ac.cn}
	\correspondingauthor{Pengfei Zhang}
	\email{zhangpengfei@ynu.edu.cn}
	\correspondingauthor{Benzhong Dai}
	\email{bzhdai@ynu.edu.cn}

	\author{Shenbang Yang}
	\affiliation{School of Physics and Astronomy, Key Laboratory of Astroparticle Physics of Yunnan Province, Yunnan University, Kunming 650091, China}
	\author[0000-0003-4895-1406]{Dahai Yan}
	\affiliation{Key Laboratory for the Structure and Evolution of Celestial Objects, Yunnan Observatory, Chinese Academy of Sciences, Kunming 650011, China}
	\author{Pengfei Zhang}
	\affiliation{School of Physics and Astronomy, Key Laboratory of Astroparticle Physics of Yunnan Province, Yunnan University, Kunming 650091, China}
	\author[0000-0001-7908-4996]{Benzhong Dai}
	\affiliation{School of Physics and Astronomy, Key Laboratory of Astroparticle Physics of Yunnan Province, Yunnan University, Kunming 650091, China}
	\author[0000-0002-5880-8497]{Li Zhang}
	\affiliation{School of Physics and Astronomy, Key Laboratory of Astroparticle Physics of Yunnan Province, Yunnan University, Kunming 650091, China}

	\begin{abstract}
		
		Blazar variability may be driven by stochastic processes.  
		On the other hand, quasi-periodic oscillation (QPO) behaviors are recently reported to be detected in \textit{Fermi}-LAT data of blazars. 
		However, the significances of these QPO signals given by traditional Fourier-like methods are still questioned.
		We analyze $\gamma$-ray light curves of the QPO blazars with two Gaussian process methods, CARMA and \textit{celerite}, 
		to examine the appropriateness of Gaussian processes for characterizing  $\gamma$-ray light curves of blazars and the existence of the reported QPOs. 
		We collect a sample of 27 blazars with possible $\gamma$-ray periodicity and generate their $\sim11$ years \textit{Fermi}-LAT light curves. 
		We apply the Gaussian process models to the $\gamma$-ray light curves, and build their intrinsic power spectral densities (PSDs).
		The results show that in general the $\gamma$-ray light curves can be characterized by CARMA and \textit{celerite} models, 
		indicating that $\gamma$-ray variabilities of blazars are essentially Gaussian processes.
		The resulting PSDs are generally the red noise shapes with slopes between $-0.6$ and $-1.7$. 
		Possible evidence for the $\gamma$-ray QPOs in PKS 0537$-$441 and PG 1553$+$113 are found in the Gaussian process modelings.
		
	\end{abstract}
	
	
	
	\section{Introduction} \label{sec:intro}
	
	Blazars are an extreme class of active galactic nuclei (AGNs) with their relativistic jets closely aligning with our line of sight, 
	whose central engines are the super massive black holes (SMBHs) located in the cores of the host galaxies. 
	Blazar emission is dominated by the Doppler-boosted and non-thermal emission of the powerful jet.
	Blazars are classified into BL Lac objects (BL Lacs) and flat spectrum radio quasars (FSRQs) according to their broad emission lines \citep[e.g.,][]{1991ApJ...374..431S,1995PASP..107..803U}. 
	FSRQs have strong emission lines, whereas BL Lacs have weak/no emission lines.
	The blazars included in the fourth catalog of AGN detected by the {\it Fermi} Large Area
		Telescope (LAT) consist of 38\% BL Lacs, 24\% FSRQs, and 38\% blazar candidates of unknown types \citep{2020ApJ...892..105A}.
		On average, FSRQs have softer spectra and stronger variabilities in GeV $\gamma$-ray energies, compared with BL Lacs  \citep{2020ApJ...892..105A}, 
		although LAT spectra of FSRQs and BL Lacs display different properties in flares \citep[e.g.,][]{2014ApJ...789..135W,2015ApJ...807...79H}.
	
	Flux variabilities of blazars have been detected at entire electromagnetic wavelengths from radio band to $\gamma$-rays.
	The variability timescales cover many orders of magnitude,
	from decades down to minutes.
	Variability analysis is a powerful tool to probe the nature of blazar \citep[e.g.,][]{2019Galax...7...28R}.
	
	Thanks to the successful operation of the \textit{Fermi}-LAT, blazar variability study has progressed significantly in the GeV $\gamma$-ray regime.
	Rapid $\gamma$-ray flare on timescale of a few minutes has been detected by LAT for the FSRQ 3C 279 \citep{2016ApJ...824L..20A}.
	The rapid flare is considered to indicate extreme jet conditions \citep[][]{2016ApJ...825L..11P,2017MNRAS.467L..16P,2020ApJ...890...56A}.
	Besides, with the LAT collecting data for more than 10 years, the long-term characteristic of the flux variability becomes attractive.
	
	In the last few years, possible quasi-periodic oscillations (QPOs) have been found in the LAT data of  more than twenty blazars  \cite[e.g.,][]{2015ApJ...813L..41A,2016ApJ...820...20S,2017ApJ...835..260Z,2017ApJ...842...10Z,2017ApJ...845...82Z,2017MNRAS.471.3036P,2018AA...615A.118S,2018NatCo...9.4599Z,2020ApJ...891..163Z,2020arXiv200200805P}. 
	The timescales of these QPOs span from one month to three years, and BL Lacs and FSRQs seem have the similar periodicity behaviors \citep[e.g.,][]{2020ApJ...891..163Z}. 
	The year-type QPOs are important for the studies on accretion disk, relativistic jet, and SMBH physics \cite[e.g.,][]{2015ApJ...813L..41A}. 
	
	The results of $\gamma$-ray QPOs \citep[e.g.,][]{2015ApJ...813L..41A,2020ApJ...891..163Z} are obtained by performing frequently used techniques such as Lomb-Scargle periodogram \cite[LSP; ][]{1976Ap&SS..39..447L,1982ApJ...263..835S} and weighted wavelet Z-transform \cite[WWZ; ][]{1996AJ....112.1709F}.
	The above two methods calculate Fourier-like power spectral density (PSD) of $\gamma$-ray light curve, and
		the QPO signal appears as a peak in the PSD.
	However, the current LAT data only cover a few cycles of the QPOs.
	Therefore, the significance for these QPOs is questionable.
	
	Some recent investigations have challenged the results of the $\gamma$-ray periodicity searchings, 
	and it is cautioned that those $\gamma$-ray QPOs may be fake signals \cite[e.g.,][]{2019MNRAS.482.1270C,2020A&A...634A.120A}. 
	Indeed, \citet{2016MNRAS.461.3145V} have pointed that the searching of periodic signals among a large number of stochastic time series should be very careful. 
	Based on the analysis with a damped random walk (DRW) model, they show that physical periodic and stochastic periodic signals are very difficult to be distinguished 
	when the time series only cover a few cycles ($\lesssim 5$ cycles). 
	\cite{2019MNRAS.482.1270C} reanalyze LAT data of 10 blazars which were reported with possible $\gamma$-ray QPOs, and they do not find any QPO signals in their sample. 
	\cite{ 2020A&A...634A.120A} analyze 9.5-year LAT data of six QPO blazars, and also do not find strong evidence for any QPOs.
	Therefore, the traditional techniques for periodicity searching in blazar variability might be misleading, and the reported QPOs should be diagnosed with different methods.
	
	An alternative approach is to fit the light curves using Gaussian processes in the time domain \citep[e.g.,][]{2009ApJ...698..895K,2010ApJ...708..927K,2010ApJ...721.1014M}.
	In this case, the PSD can be calculated from the inferred autocovariance function \citep{2014ApJ...788...33K}.
	In this field, the representative method is the continuous-time autoregressive moving average (CARMA) models developed by \citet{2014ApJ...788...33K}.
	Actually, the DRW model  is the case of CARMA(1,0). 
	CARMA models are flexible to capture the features of flux variability and to produce more accurate PSD.
	The CARMA method has been already applied to LAT data of blazars \citep[e.g.,][]{2018ApJ...863..175G,2019ApJ...885...12R}.
	Comparably, \cite{2017AJ....154..220F} develop an other fast and flexible Gaussian process model, \textit{celerite}, for estimating
	the variability features of a light curve and its PSD. 
	The capability of  \textit{celerite} for characterizing the variability of AGN is unknown, and it requires more investigations to find out its potentiality in the AGN variability studies.
	
	In this paper,  
	we carefully study the $\gamma$-ray light curves of 27 $Fermi$ blazars which have possible QPO features, using the different Gaussian process methods: CARMA and \textit{celerite}.
	We aim to primarily figure out two points: 
	(i) whether the $\gamma$-ray light curves can be fitted well by the Gaussian processes or not;
	(ii) whether the reported QPO signals still appear in the derived PSDs or not.
	Additionally, we also test the efficiency of \textit{celerite} model for the AGN variability studies.
	The format of this paper is as follows.
	In Section \ref{sec:carma}, we briefly introduce the basic concepts of CARMA and \textit{celerite}. 
	In Section \ref{sec:data}, we briefly describe the analysis procedure of $Fermi$ data covering $\sim 11$ years. 
	In Section \ref{sec:result} we give the modeling results of the light curves with CARMA and \textit{celerite}. 
	Finally, we discuss the results in Section \ref{sec:discussion}. 
	
	\section{CARMA and celerite models} \label{sec:carma}
	
	If a light curve $y(t)$ is considered as a zero-mean CARMA($a,b$) process, it can be obtained by solve the stochastic differential equation \citep{2014ApJ...788...33K}:
	
	\begin{eqnarray} \label{equ:car} 
	\alpha_a\frac{d^ay(t)}{dt^a}+\alpha_{a-1}\frac{d^{a-1}y(t)}{dt^{a-1}}+\cdots+\alpha_0y(t) \nonumber \\
	= \beta_b\frac{d^b\epsilon(t)}{dt^b}+\beta_{b-1}\frac{d^{b-1}\epsilon(t)}{dt^{b-1}}+\cdots+\beta_0\epsilon(t),
	\end{eqnarray}
	
	where $\epsilon(t)$ is a white-noise process with zero mean and variance $\sigma^2$, and $\alpha_a$ and $\beta_0$ are defined to equal 1. 
	The PSD of the stationary CARMA($a,b$) process can be written as 
	\begin{equation} \label{equ:car_psd}
	P(f)=\sigma^2\frac{\left|\sum^b_{j=0}\beta_j(2\pi if)^j \right|^2}{\left| \sum^a_{k=0}\alpha_k(2\pi if)^k \right|^2}.
	\end{equation}
	In practice, the parameters $\sigma^2$, $\alpha_{0},\cdots,\alpha_{a-1}$, $\beta_{1},\cdots$, and $\beta_{b}$ can be obtained by fitting $y(t)$ to a observed light curve 
	with determined $a$ and $b$, and then it is convenient to calculate the PSD using equation (\ref{equ:car_psd}). 
	Any characteristic timesacle including QPO could manifest a structure on the PSD. 
	For more details about CARMA, the reader is referred to \citet{2014ApJ...788...33K}. 
	The software we used in this paper is \texttt{carma\_pack}\footnote{\url{https://github.com/brandonckelly/carma\_pack}} released by \citet{2014ApJ...788...33K}.
	
	\textit{celerite} \citep{2017AJ....154..220F} is also a Gaussian process method for modeling light curve, and shares many features with CARMA. 
	According to its original kernel function (i.e., covariance function), it can be used to calculate any CARMA model theoretically. 
	The official software \texttt{celerite}\footnote{\url{https://github.com/dfm/celerite}} is designed to be restrictive 
	and it requires a specific kernel function chosen or provided by the user. 
	In other words, users are free to analyze their time series with stochastic or physical models. 
	Nevertheless, we adopt the stochastically driven damped \textit{simple harmonic oscillator} (SHO) as the model for our condition, 
	following the recommendation of the developers. 
	A SHO system has the differential equation:
	\begin{equation}\label{equ:cele}
	\left[\frac{d^2}{dt^2}+\frac{\omega_0}{Q}\frac{d}{dt}+\omega_0^{2}\right]y(t)=\epsilon(t),
	\end{equation}
	where $\omega_0$ is the frequency of the undamped oscillator, $Q$ is the
	quality factor of the oscillator, and $\epsilon(t)$ is assumed to be white noise. Then the PSD can be expressed as 
	\begin{equation}\label{equ:cele_psd}
	P(\omega)=\sqrt{\frac{2}{\pi}}\frac{P_0\omega_0^4}{\left(\omega^2-\omega _0^2\right)^2+\omega_0^2\omega^2/Q^2},
	\end{equation}
	where $P_0$ is proportional to the power at $\omega=\omega_0$. 
	As CARMA does, a specific model for a time series can be built up with a mixture of determined number of SHO terms. 
	The number of the oscillators should be determined using model selection criteria. 
	
	\section{Data Analysis} \label{sec:data}
	
	We collect 27 blazars with possible $\gamma$-ray QPOs, including 17 BL Lacs and 10 FSRQs. 
	The reported period and relevant information of each blazar are demonstrated in Table \ref{tab:src_info}. 
	
	We use the \textit{Fermi}-LAT photon data of these sources covering $\sim 11$ years (from 2008 August 4 to 2019 September 5). 
	We keep only SOURCE class events (\texttt{evclass=128}) and the three event types (\texttt{evtype=3}). 
	The region of interest (ROI) for each source is a circle centered at the corresponding position with $15\degr$  radius. 
	To avoid the contamination from the Earth's limb, the maximum zenith angle is set to be 90\degr. 
	The good time intervals are selected with the recommended filter expression \texttt{(DATA\_QUAL>0)\&\&(LAT\_CONFIG==1)}. 
		Additionally, to avoid the impact from the Sun emission, we exclude the time intervals when the distances from the Sun directions to the sources are less than $15\arcdeg$, 
		with the expression \texttt{angsep(src\_ra, src\_dec, RA\_SUN, DEC\_SUN)>15}.  
	We use the script \texttt{make4FGLxml} and the source catalog \texttt{gll\_psc\_v19} to generate our model file for the likelihood analysis. 
	The Galactic and extragalactic diffuse and isotropic emission models are \texttt{gll\_iem\_v07} and \texttt{iso\_P8R3\_SOURCE\_V2\_v1}, respectively.
	The instrument response function  \texttt{P8R3\_SOURCE\_V2} and the unbinned likelihood analysis method are used.
	The whole procedure is following the official unbinned likelihood tutorial\footnote{\url{https://fermi.gsfc.nasa.gov/ssc/data/analysis/scitools/likelihood\_tutorial.html}}, and is performed in the Fermitools-conda (version 1.0.10) environment. 
	\startlongtable
	\begin{deluxetable*}{lcccccc}
		\tablecaption{Information of the 27 blazars. The reported periods are collected from the references in last column. \label{tab:src_info}}
		\tablehead{
			\colhead{4FGL Name} & \colhead{RA} & \colhead{DEC} & \colhead{Identification } & \colhead{Type} & \colhead{Reported Period} & \colhead{Ref.} \\
			\colhead{} & \colhead{} & \colhead{} & \colhead{} & \colhead{} & \colhead{(yr)} & \colhead{}
		}
		\startdata
		4FGL J0043.8$+$3425	&	10.9717	&	34.4316	&	GB6 J0043$+$3426	&	FSRQ	&	1.8	&	(1)	\\
		4FGL J0102.8$+$5824	&	15.701	&	58.4092	&	TXS 0059$+$581	&	FSRQ	&	2.1	&	(1)	\\
		4FGL J0210.7$-$5101	&	32.6946	&	$-$51.0218	&	PKS 0208$-$512	&	FSRQ	&	2.6	&	(1)	\\
		4FGL J0211.2$+$1051	&	32.8091	&	10.8569	&	MG1 J021114$+$1051	&	BLL	&	1.7	&	(1)	\\
		4FGL J0252.8$-$2219	&	43.2007	&	$-$22.3203	&	PKS 0250$-$225	&	FSRQ	&	1.2	&	(1)	\\
		4FGL J0303.4$-$2407	&	45.8625	&	$-$24.1225	&	PKS 0301$-$243	&	BLL	&	2.1	&	(2)	\\
		4FGL J0428.6$-$3756	&	67.173	&	$-$37.9403	&	PKS 0426$-$380	&	BLL	&	3.3	&	(3)	\\
		4FGL J0449.4$-$4350	&	72.3582	&	$-$43.835	&	PKS 0447$-$439	&	BLL	&	2.5	&	(1)	\\
		4FGL J0457.0$-$2324	&	74.2608	&	$-$23.4149	&	PKS 0454$-$234	&	FSRQ	&	2.6	&	(1)	\\
		4FGL J0501.2$-$0158	&	75.3023	&	$-$1.9749	&	S3 0458$-$02	&	FSRQ	&	1.7	&	(1)	\\
		4FGL J0521.7$+$2112	&	80.4445	&	21.2131	&	TXS 0518$+$211	&	BLL	&	2.8	&	(1)	\\
		4FGL J0538.8$-$4405	&	84.7089	&	$-$44.0862	&	PKS 0537$-$441	&	BLL	&	0.77	&	(4)	\\
		4FGL J0721.9$+$7120	&	110.4882	&	71.3405	&	S5 0716$+$714	&	BLL	&	0.95/2.8	&	(5)/(1)	\\
		4FGL J0808.2$-$0751	&	122.065	&	$-$7.8556	&	PKS 0805$-$077	&	FSRQ	&	1.8	&	(5)	\\
		4FGL J0811.4$+$0146	&	122.861	&	1.7756	&	OJ 014	&	BLL	&	4.3	&	(1)	\\
		4FGL J0818.2$+$4222	&	124.5572	&	42.3819	&	S4 0814$+$42	&	BLL	&	2.2	&	(1)	\\
		4FGL J1058.4$+$0133	&	164.624	&	1.5641	&	4C $+$01.28	&	BLL	&	1.22	&	(5)	\\
		4FGL J1146.9$+$3958	&	176.7405	&	39.9775	&	S4 1144$+$40	&	FSRQ	&	3.3	&	(1)	\\
		4FGL J1248.3$+$5820	&	192.0844	&	58.3432	&	PG 1246$+$586	&	BLL	&	2	&	(1)	\\
		4FGL J1303.0$+$2434	&	195.7571	&	24.5821	&	MG2 J130304$+$2434	&	BLL	&	2	&	(1)	\\
		4FGL J1555.7$+$1111	&	238.9313	&	11.1884	&	PG 1553$+$113	&	BLL	&	2.2	&	(6)	\\
		4FGL J1649.4$+$5235	&	252.3637	&	52.5901	&	87GB 164812.2$+$524023	&	BLL	&	2.7	&	(1)	\\
		4FGL J1903.2$+$5540	&	285.8077	&	55.6773	&	TXS 1902$+$556	&	BLL	&	3.8	&	(1)	\\
		4FGL J2056.2$-$4714	&	314.0715	&	$-$47.2369	&	PKS 2052$-$477	&	FSRQ	&	1.75	&	(5)	\\
		4FGL J2158.8$-$3013	&	329.7141	&	$-$30.2251	&	PKS 2155$-$304	&	BLL	&	1.76	&	(7);(8)	\\
		4FGL J2202.7$+$4216	&	330.6946	&	42.2821	&	BL Lacertae	&	BLL	&	1.9	&	(5);(8)	\\
		4FGL J2258.1$-$2759	&	344.5288	&	$-$27.9843	&	PKS 2255$-$282	&	FSRQ	&	1.3	&	(1)	\\
		\enddata
		\tablerefs{The number in last column ( Ref.) denote: (1) \citet{2020arXiv200200805P}, (2) \citet{2017ApJ...845...82Z}, (3) \citet{2017ApJ...842...10Z}, (4) \citet{2016ApJ...820...20S}, (5) \citet{2017MNRAS.471.3036P}, (6) \citet{2015ApJ...813L..41A}, (7) \citet{2017ApJ...835..260Z}, (8) \citet{2018AA...615A.118S}}
	\end{deluxetable*}
	
	To produce the light curve, we first select the data between 100 MeV and 300 GeV in the time span given above, and perform the binned likelihood analysis to get the best-fit parameters. 
	Then we set these best-fit parameters as initial parameters in the input model file, and use the unbinned likelihood method to generate the light curves. 
	All the parameters of sources at the positions greater than $10\degr$ from the center of ROI are fixed.  
	The normalizations of the sources in the region of $\leqslant10\degr$ are set to be free.
	For each source, two light curves are built: one 30-day binning light curve and one 14-day binning light curve.

	
	\section{Results} \label{sec:result}
	
	At first, we would like to stress that the use of the Gaussian methods is limited for faint sources, 
	           since one needs to include also the non-robust detections in the analysis for having a sufficient number of the data points required for these models. 
	           Consequently, the results given for such sources should be treated with caution and tested by means of the alternative models for which a relatively few number of the detections (only the robust ones) are required. 
	
	
	Therefore, we exclude the data points with Test Statistic (TS) value  of $<25$ or the predicted photon number of $N_{\rm pred} < 10$ in the light curves 
	in order to get reliable results \citep[see, e.g., ][]{2013MNRAS.436.1530R,2018MNRAS.480..407K,2020ApJS..247...27K}.
	Unfortunately, there are 3 sources, GB6 J0043+3426, MG2 J130304+2434, and 87GB 164812.2+524023, whose light curves cannot be selected by the criterion given above, because they are too faint to meet the criterion.
	We then apply a loose exclusion criterion of TS $<4$ to the light curves of the 3 faint sources.

	Note that, as shown in \citet{2014ApJ...788...33K} and \citet{2017AJ....154..220F}, CARMA and \textit{celerite} models have the capability of handling with the irregular sampling, 
	and they both consider the errors of the data points in the modeling.
  
	We perform CARMA model selections to the light curves with an extended parameter space of $a=1,\cdots,7,b=0,\cdots,a-1$.
	In modeling a light curve, Markov Chain Monte
	Carlo (MCMC) technique is used to sample the parameters, and a bayesian criterion deviance information criterion (DIC) is calculated \citep[][]{2014ApJ...788...33K}.
	The model with the minimum DIC is considered as the best one. 
	
	For \textit{celerite}, a more complicated model is needed for fitting the real data. 
	Therefore, DRW is added to the SHO model, as the DRW is used to characterize the aperiodic component in variability \cite[e.g.,][]{2018ApJ...859L..12L,2020ApJ...895..122C}. 
	Namely, the model can be written as DRW+SHO$\times n$. 
	In this work, $n$ is set to be $\leqslant4$. We perform the corrected \textit{Akaike information criterion} (AIC) to do the model selection. 
	The AIC is defined as 
	\begin{equation} \label{eq:aic}
	{\rm AICc} = 2k-2\log\mathcal{L}+\frac{2k(k+1)}{n-k-1},
	\end{equation}
	where $k$ is the number of parameters, $n$ is the number of data points of light curves, $\mathcal{L}$ is the maximum likelihood. 
	We improve the fitting procedure in \textit{celerite} by adopting the MCMC sampler \texttt{emcee}\footnote{\url{https://github.com/dfm/emcee} } \citep{2013PASP..125..306F} .
	We run 100 optimizations with random starting values of parameters to avoid the effect caused by the instability of algorithm \texttt{L-BFGS-B}. 
	We calculate maximum likelihood for each optimization, and the maximum value among the output is used to calculate the AIC.  
	We perform this procedure on each model, and the best model is the one with the minimum AIC. 
	
	In our analysis, we run MCMC sampler for 50000 iterations in CARMA model. 
	The first 25000 iterations are taken as burn-in sampling, which is not involved in the posterior analysis. 
	In \textit{celerite} model, we run \texttt{emcee} sampler using 32 parallel walkers for 10000 steps as burn-in and 20000 steps for MCMC sampling. 
	
	\subsection{Modeling Light Curves}\label{subsec:modeling_lc}
	
\begin{figure}[h]
	\includegraphics[width=0.5\textwidth]{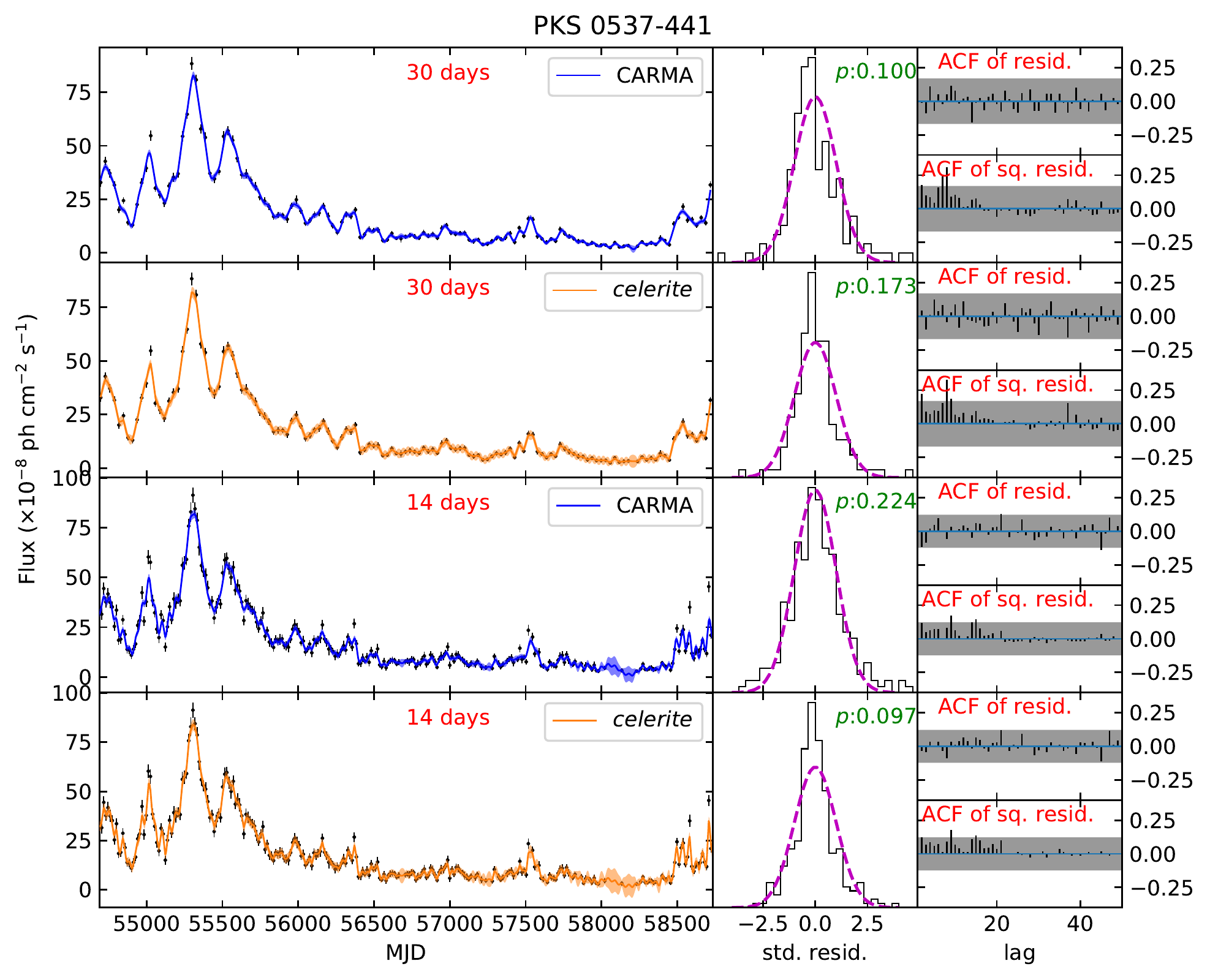}

	\caption{Fitting result for PKS 0537-441, as an example of the 24 bright sources. 
		For each source, we give the LAT light curves with two time bins (black points) and the modeled light curves in left column.
		The time bins are labeled in the figures. Blue and bright orange lines represent CARMA and \textit{celerite} modeling results, respectively. 
		In central column, we show the probability density of fitting standardized residuals (black solid line) and the expected normal distribution (the magenta line).
		The $p$-values of KS tests for the standardized residuals are also reported. The ACF of residuals and squared residuals are showed in right column where the gray region is the $95\%$ confidence limit of the white noise. The complete figure set (24 images) is available in the online journal.  \label{fig:fitlc}}
\end{figure}

\begin{figure*}
	\centering
	\includegraphics[width=0.495\textwidth]{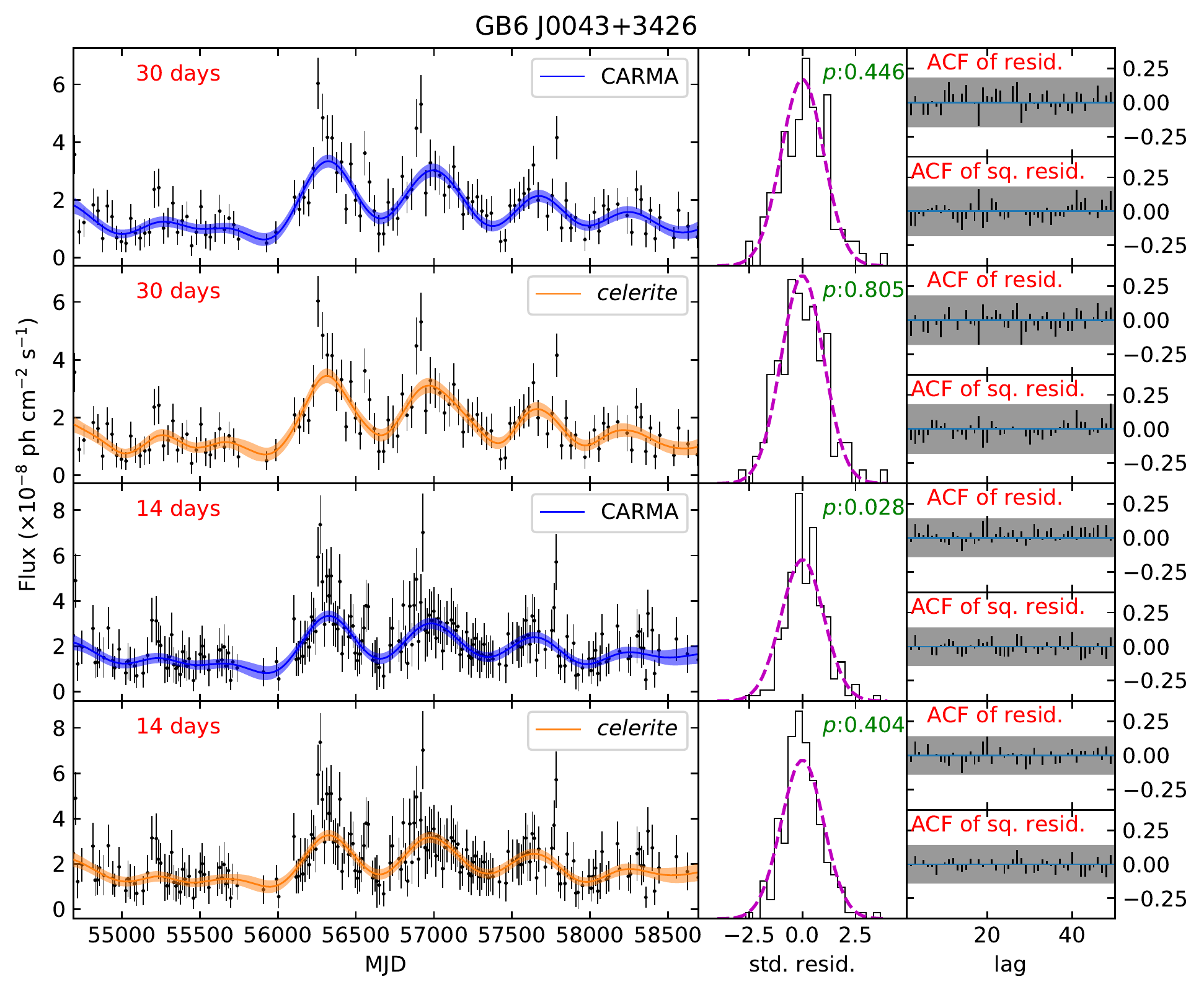}
	\includegraphics[width=0.495\textwidth]{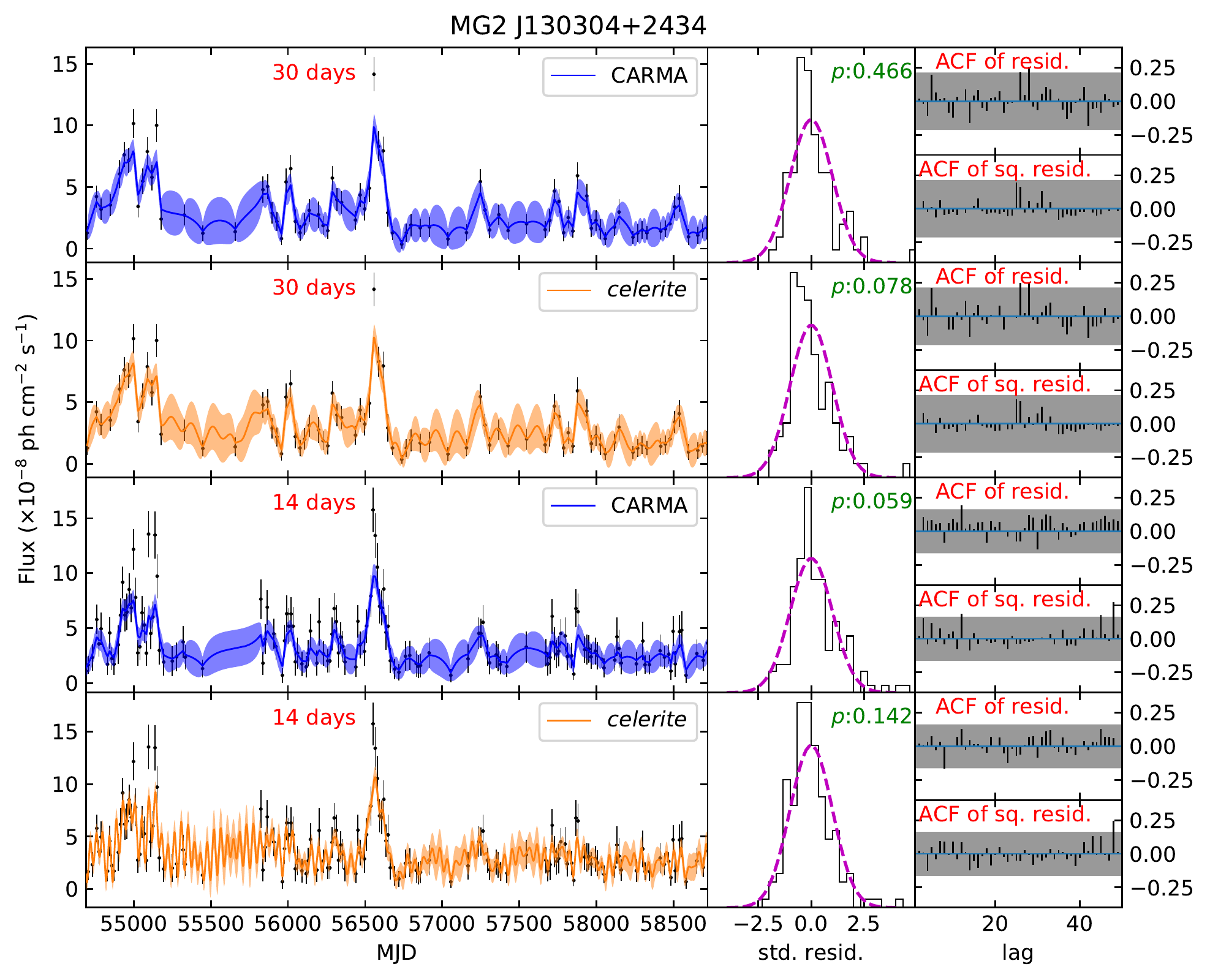}
	\includegraphics[width=0.495\textwidth]{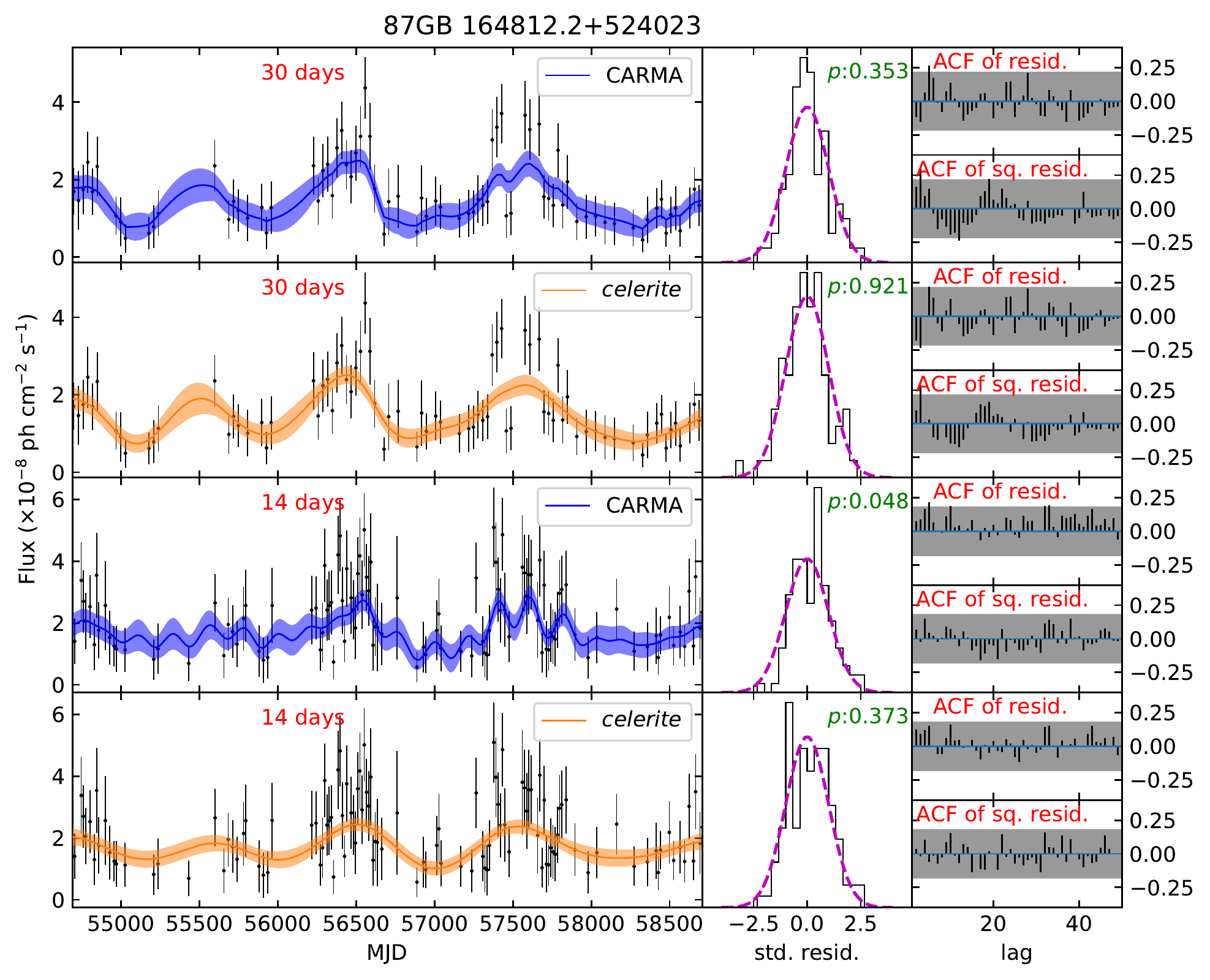}
	\caption{Fitting results for the 3 faint sources. The symbols and lines are the same as that in Figure \ref{fig:fitlc}. \label{fig:fitlc_3}}
\end{figure*}

	The $\gamma$-ray light curves in the 0.1-300 GeV energy range of the 27 $\gamma$-ray blazars are respectively modeled by CARMA and  \textit{celerite}.
	The LAT light curves and modeled light curves are shown in Figure \ref{fig:fitlc} (for the 24 bright sources) and Figure \ref{fig:fitlc_3} (for the 3 faint sources).
	The modeling results are given in Table \ref{tab:modeling_res}.
	
	To assess the goodness of the fitting with Gaussian models, 
	we perform a standardized residual analysis.
	In the Kalman filter approach that is used in CARMA, the standardized residuals, $\bm{\chi}$, can be calculated as 
	
	\begin{equation}
	\chi_i = \frac{y_i - \hat{E}_i}{\sqrt{Var(\hat{E}_i)}},
	\end{equation} 
	where $y_i$ is the observation, $\hat{E}_i$ is the expectation from the Kalman filtering, 
	$Var(\hat{E}_i)$ is the variance of $\hat{E}_i$. 
	
	To link Gaussian process of \textit{celerite} to moving averages, the observed data can be expressed as 
	\begin{equation}
	\textbf{y*}= {\rm chol}\left(  \textbf{K} \right)^{\top}  \textbf{w},  
	\end{equation}
	where $\textbf{y*}$ is zero-mean vector of observation data, 
	${\rm chol}\left(  \textbf{K} \right)$ is the upper triangular Cholesky factorization of the covariance matrix $\textbf{K}$, 
	and $\textbf{w}$ is the white noise which can be considered as the standardized residuals described above. 
	Thus $\textbf{w}$ can be calculated as
	\begin{equation}
	\textbf{w}= {\rm chol}\left(  \textbf{K} \right)^{-\top} \textbf{y*} .
	\end{equation}
	The expected $\hat{E}_i$ and $\textbf{K}$ are appropriately obtained using the maximum-likelihood estimates.
	
	If the model is correct, the $\bm{\chi}$ or $\textbf{w}$, which is referred as standardized residuals here, should be approximately a normal distribution with mean zero and standard deviation of one. 
	The standardized residuals should follow a Gaussian white-noise sequence, which can be assessed through the auto-correlation function (ACF) of the standardized residuals.
	
	With the modeling results, we calculate the standardized residuals and analyze their probability densities. 
	We first examine the deviation of the distribution of the standardized residuals from a normal distribution using the Kolmogorov-Smirnov (KS) test. 
	The $p$-values of the tests are given in Table \ref{tab:modeling_res}. 
	We then calculate the ACF of the residual and squared residual sequences to assess their behaviors (see Figure \ref{fig:fitlc}). 
	
	Looking at the distribution of the standardized residuals, one can find that the residuals of most of the fits are consistent with the expected normal distribution with $p>0.05$.
	The 30-day light curves have smaller errors and their rapid flares are easier to be smoothed, compared with the 14-day light curves.
	In general, both CARMA and \textit{celerite} can well describe the light curves of the two time bins.
	The ACFs of the residuals and squared residuals are almost inside the $95\%$ confidence limits, 
	which indicates that the models have captured the correlation structures and no significant nonlinear behaviors appear in the time series. 
	
	However, we also notice that the distribution of the standardized residuals of several sources (e.g., PKS 0250$-$225, PKS 0426$-$380, PKS 0454$-$234, S3 0458$-$02, S5 0716$+$714 and PKS 0805$-$077) deviate from the expected normal distribution.
	The distributions of the standardized residuals of these sources are concentrated around zero, suggestive of overfitting. 
	The overfitting is likely due to the errors of the data.
	
	Generally speaking, a simple DRW (or Ornstein-Uhlenbeck process) model fails to model the light curves, 
	and high-order CARMA and \textit{celerite} models are needed to characterize the $\gamma$-ray variabilities (Table~\ref{tab:modeling_res}).
	This is consistent with the results of previous works \citep{2014ApJ...786..143S,2018ApJ...863..175G,2019ApJ...885...12R} obtained from different samples.
	
	It is also worth noting that both CARMA and \textit{celerite} models cannot characterize large amplitude flares (see the results of PKS 0301-243), 
	despite that this failure is statistically unmeaning.
	
	No significant difference is found between the modeling results of CARMA and \textit{celerite}.

\figsetstart
\figsetnum{1}
\figsettitle{Fitting results for the 24 bright sources.}
\figsetgrpstart
\figsetgrpnum{1.11}
\figsetgrptitle{PKS 0537-441}
\figsetplot{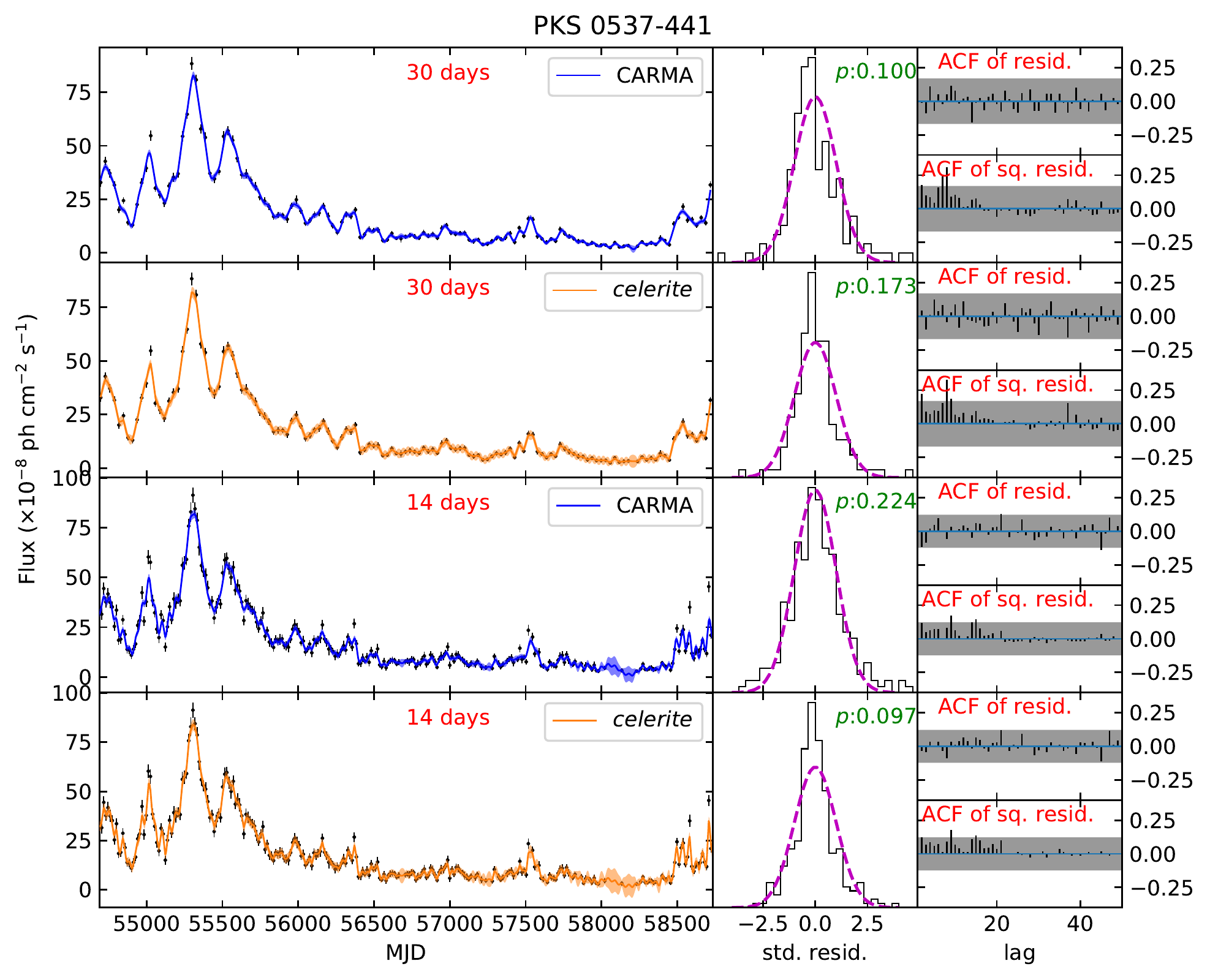}
\figsetgrpnote{Fitting results for PKS 0537-441.}
\figsetgrpend
\figsetend

	\subsection{Building PSDs}\label{subsec:buidingPSD}
	
	The PSD can be constructed by calculating Equations (\ref{equ:car_psd}) and (\ref{equ:cele_psd}) with the modeling results.
	The resulting PSDs for the 24 bright sources are shown in Figure \ref{fig:allpsd}, and the PSDs for the three faint sources are shown in Figure \ref{fig:psd_3}.
	We  also give the noise level  caused by
	measurement errors, which is estimated by (assuming Gaussian errors)
	\begin{equation} \label{eq:noise}
	P_{\rm noise}=2\times\Delta t\times\overline{\sigma_{\rm err}^{2}},
	\end{equation} 
	where $\Delta \rm t$ is the median of sampling time and $\sigma_{\rm err}$ is the measurement error. 
	It should be noted that such a noise has been deducted from each PSD.
	
	We estimate the slope of each PSD sample by fitting it with a linear function in log-log space. 
	The structure at low frequencies ($\sim10^{-3} \rm~ day^{-1}$) is not taken into account in the fitting.
	The average slopes and the corresponding confidence intervals are then calculated. 
	The results are given in Table \ref{tab:modeling_res}.
	
\begin{figure*}
	\centering
	\includegraphics[width=0.5\textwidth]{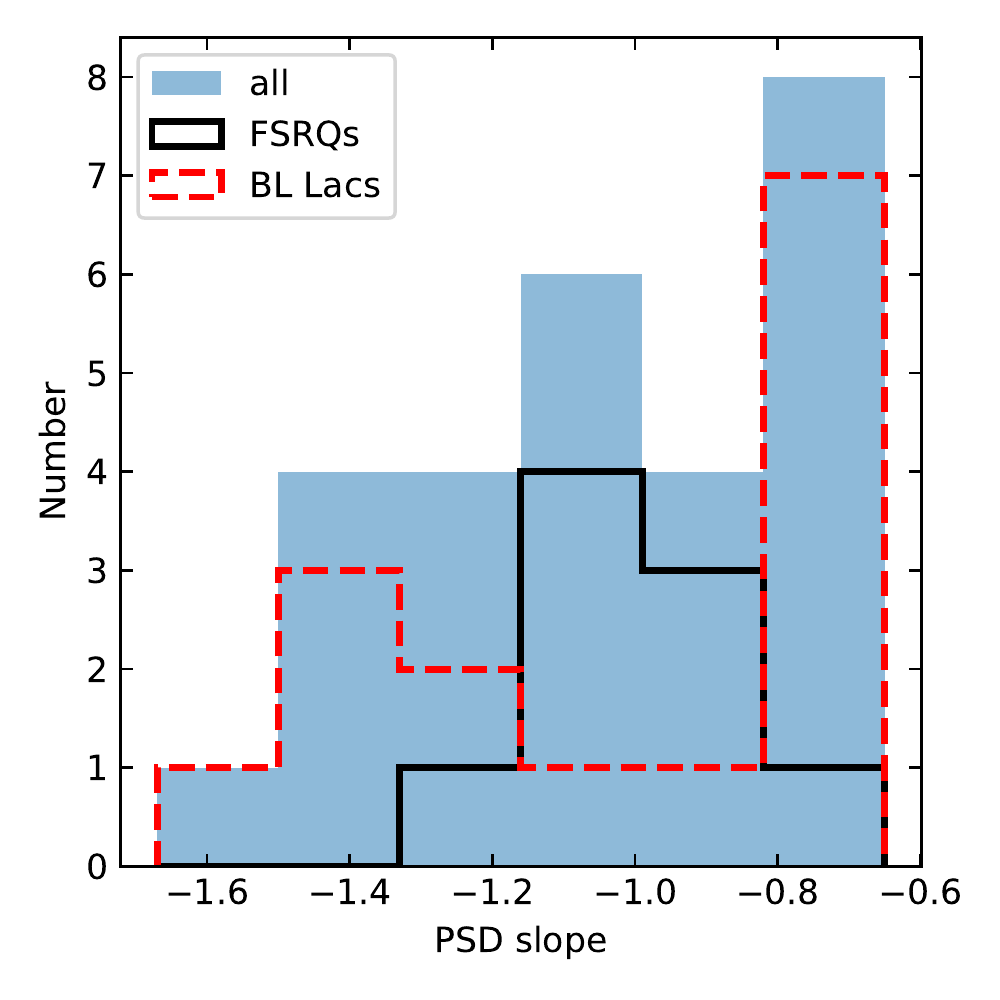}
	\caption{Distributions of the slopes of PSDs produced by CARMA. The blue bars are the whole distribution of those for the 24 bright sources. 
	           The black solid line is the distribution of those for FSRQs, while the red dotted line is the one for BL Lacs.  \label{fig:psd_slopes}}
\end{figure*}

	From~Table \ref{tab:modeling_res}, one can find that the PSDs produced by CARMA and \textit{celerite} are consistent for all sources.
	Considering the uncerntainties, there is no divergence between the PSDs of the light curves in different time bins.
	For the PSDs above 100 MeV, the slope is between $-0.6$ and $-1.7$, as shown in Figure~\ref{fig:psd_slopes}.
	FSRQs and BL Lacs have similar distributions of the PSD slopes (Figure~\ref{fig:psd_slopes}).
	It should be stressed that the errors on the slopes are large for all sources. 
		
\begin{figure*}
	\gridline{\fig{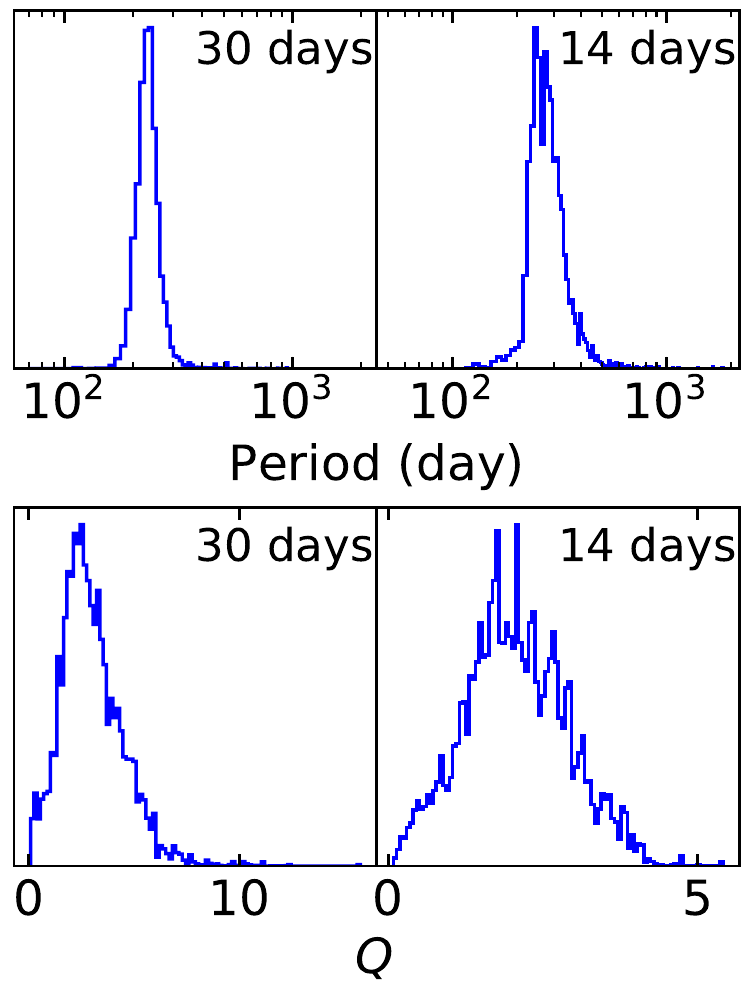}{0.38\textwidth}{(a)}
		 \fig{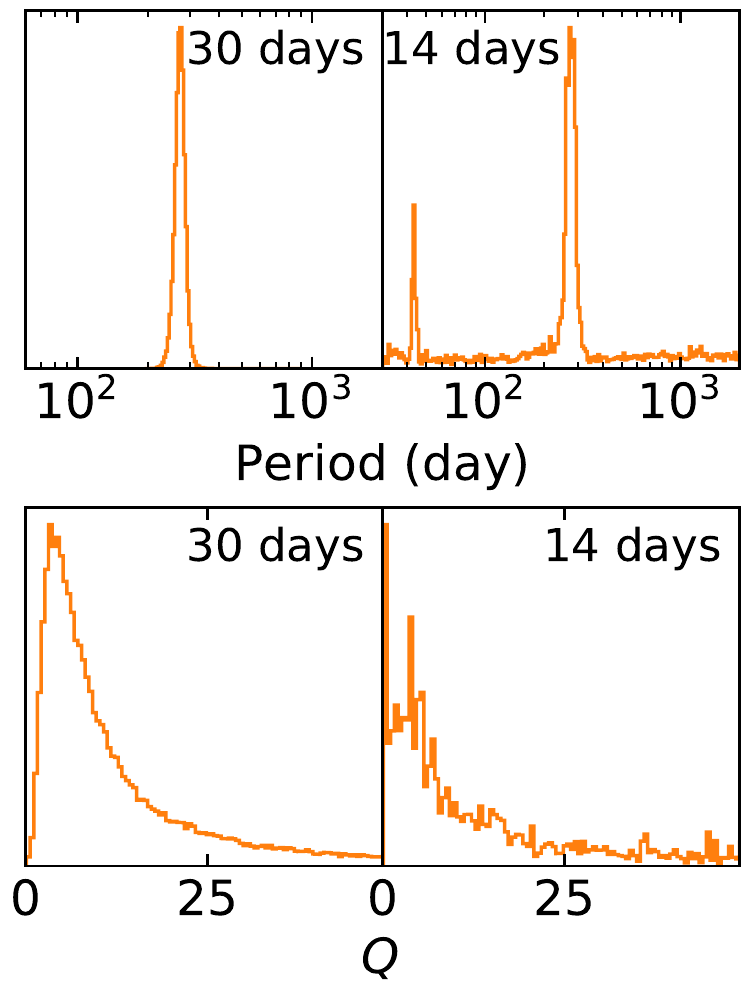}{0.38\textwidth}{(b)}}
	\caption{Marginalized posterior probability densities of period and the corresponding quality factor $Q$ for PKS 0537-441. (a) the distributions from CARMA (blue), (b) the distributions from \textit{celerite} (bright orange).\label{fig:pks0537post}}
\end{figure*}

	No QPO feature is found in all \textit{celerite} and CARMA PSDs, except for PKS 0537$-$441 and PG 1553$+$113.
	We further analyze the posterior distributions of the QPO features in PKS 0537$-$441 and PG 1553$+$113 (Table~\ref{tab:post} and Figures~\ref{fig:pks0537post}--\ref{fig:pg1553post}).
	According to the Nyquist criterion, the lower limit of the posterior distribution of period is 60 days when the time bin is 30 days, and is 28 days when the time bin is 14 days. 
	The calculation of the uncertainties on the period considers the whole posterior probability density distribution.
	
	The period of PKS 0537$-$441 is constrained well at $\sim 233^{+27}_{-23}$ days by CARMA in the 30-day light curve (Table \ref{tab:post} and Figure \ref{fig:pks0537post}).
	Similarly, the period in the 14-day light curve is at $\sim 280^{+63}_{-40}$ days.
	The two results are consistent within their errors.
	\textit{celerite} captures a QPO signal at $\sim 273^{+13}_{-14}$ days in the 30-day light curve.
	For the 14-day light curve, a sign of a period of $\sim 270$ days appears (Table \ref{tab:post} and Figure \ref{fig:pks0537post}).
	
	For PG 1553$+$113, a period of $\sim840^{+338}_{-164}$ days is found by CARMA in the 30-day binning light curve.
	Note that the uncertainties on the period are large.
	No QPO is captured in the 14-day light curve by CARMA (Figure~\ref{fig:pg1553post} and Table~\ref{tab:post}).
	In \textit{celerite} model, the probability densities of the period obtained from two light curves both show a peak at $\sim 800$ days (Figure~\ref{fig:pg1553post}), 
	however, in both cases the period is not constrained (Table \ref{tab:post}). 

	\subsection{Estimating the Significance of the QPOs in PKS 0537$-$441 and PG 1553$+$113}
	
	Generally, quality factor ($Q$) in CARMA and $celerite$ models can be used to quantify the coherence of a QPO \citep{2014ApJ...788...33K,2017AJ....154..220F}, 
	which is defined as the ratio of the centroid frequency of the QPO to its width.
	\cite{2017AJ....154..220F} showed that the QPO feature only appears in the PSD when $Q > 0.5$.
	We analyze the posterior distributions of the $Q$ values (Figures \ref{fig:pks0537post} and \ref{fig:pg1553post}), 
	and the details are given in Table \ref{tab:post}. 
	The $Q$ values for the QPO of PKS 0537$-$441 are constrained in CARMA model, and it is $\sim 2$. 
	$Q$ is poorly constrained in $celerite$ modelings for both the two light curves of PKS 0537$-$441. 
	For PG 1553$+$113, CARMA gives $Q\sim1.4$, but with large errors;
	while in $celerite$ model, $Q$ is not constrained.

	With the $Q$ values obtained in the MCMC sampling, 
	we can evaluate the significance of the QPO signal \citep{2014ApJ...788...33K}.
	Following \citet{2017AJ....154..220F}, we consider that the QPO with $Q > 0.5$ is true.
	For the period that we are interested in, the peak in the period posterior probability distribution is fitted by a Gaussian function.
	The range covered by 99.9999\% of the Gaussian distribution is considered.
	In this range, the number of the QPO with $Q > 0.5$, $N_{\rm true}$, is counted.
	We then can evaluate the significance of the QPO of interest, and the significance=$N_{\rm true}$/$N_{\rm tot}$.
	When $N_{\rm tot}$ is the total number of $Q$ in whole analysis, we obtain the global significance; 
	when $N_{\rm tot}$ is just the total number of $Q$ in the period range mentioned above, we obtain the so-called local significance, 
	i.e., the significance of the detected period at that position \citep[e.g.,][]{2020A&A...634A.120A}. 

	The significances for the QPOs in PKS 0537$-$441 and PG 1553$+$113 are listed in Table \ref{tab:post}. 
	The global significance of the period in PKS 0537$-$441 given by CARMA is $>95\%$. It is $>99\%$ given by the $celerite$ model with the 30-day binning light curve.
	The corresponding local significance is slight higher than the global significance.
	However,  the $celerite$ model gives a low global significance ($\sim 50\%$) and a higher local significance ($>90$\%) for the period with the 14-day binning light curve.
	For PG 1553+113, the global significance of the period given by CARMA model with the 30-day binning light curve is $\sim 90\%$ and the local significance is $\sim 95\%$, 
	while no QPO feature is found in the 14-day binning light curve. 
	The global significances of the periods given by $celerite$ model with the two light curves are only $\sim 50\%$, and the local significances are higher ( $>95$\%). 
	
	
	\section{Discussions and Conclusions} \label{sec:discussion}
	
	Some blazars' $\gamma$-ray light curves continuously monitored by \textit{Fermi}-LAT are mainly behaving as stochastic processes \citep[e.g.,][]{2014ApJ...786..143S,2019ApJ...885...12R}, 
	which can be explained by the internal shock model or turbulence in the jet \citep[e. g.,][]{2010arXiv1006.5364R,2014ApJ...780...87M}. 
	However, the year-type periodic behaviors in $\gamma$-ray band are recently claimed for blazars. 
	If the periodicities are confirmed, these blazars will be an interesting sample for the investigation of accretion physics \citep[e.g.,][]{2012MNRAS.423.3083M}, 
	jet geometry \citep[e.g.,][]{2004ApJ...615L...5R}, 
	and the gravitational wave induced by the SMBH binary \citep[e.g.,][]{2006MmSAI..77..733K,2007Ap&SS.309..271R,2019A&ARv..27....5B}. 
	However, the $\gamma$-ray QPOs of blazars are questioned \citep[e.g.,][]{2019MNRAS.482.1270C,2020A&A...634A.120A}.
	
	Stochastic models have been developed to characterize astronomical variability \citep[e.g.,][]{2013ApJ...765..106Z,2014ApJ...788...33K,2017AJ....154..220F,2018MNRAS.476L..55L}.
	This kind of models assumes that the light curve is a
	realization of a Gaussian process.
	CARMA and \textit{celerite} are two popular stochastic models for modeling astronomical variability.
	CARMA has been applied to $\gamma$-ray variabilities of blazars \citep{2014ApJ...786..143S,2018ApJ...863..175G,2019ApJ...885...12R}\footnote{\citet{2014ApJ...786..143S} used a model of Ornstein-Uhlenbeck process which is identical to a particular CARMA model.}.
	\textit{celerite} is a new tool in AGN variability study.
	
	\citet{2014ApJ...786..143S} applied their stochastic models to the four-year  \textit{Fermi}-LAT light curves of 13 blazars, and found that the $\gamma$-ray activity of the blazar is consistent with stochastic processes.
	\citet{2019ApJ...885...12R} modeled the 9.5-year \textit{Fermi}-LAT light curves of the same 13 blazars with CARMA, and claimed that CARMA(2,1) models provide adequate
	descriptions of the variability.
	\citet{2020ApJ...895..122C} investigated the $\gamma$-ray QPOs in PG 1553$+$113 and PKS 2155$-$304 with Gaussian process, and confirmed the QPO in 1553$+$113.
	
	So far, possible year-type $\gamma$-ray QPOs are reported in 27 blazars.
	We analyze the 11-year \textit{Fermi}-LAT data of the 27 blazars and build their light curves.
	CARMA and \textit{celerite} are applied to these light curves.
	General speaking, the $\gamma$-ray light curves can be well characterized by high-order CARMA and \textit{celerite} models, 
	which suggests that the $\gamma$-ray activity of the blazars is Gaussian process essentially.
	The overfitting problem in several blazars can be resolved by modeling the logarithmic flux (see Appendix \ref{append:a}).
	
	PSDs are produced from the modeling results.
	The PSDs recovered by CARMA and \textit{celerite} are consistent for the 27 sources with the slopes from $-0.6$ to $-1.7$.
	For all sources, there is no significant difference between the PSDs of the light curves in different time bins.
	However, it is noted that the uncertainties on the PSDs are large.
	
	QPO feature only appears in the PSDs of PKS 0537$-$441 and PG 1553$+$113.
	Further posterior analyses are performed to examine the two periodicities.
	A period of $\sim250$ days for PKS 0537$-$441 is well constrained in CARMA model.
	The modeling result of \textit{celerite} gives a period of $\sim280$ days.
	It is noted that \textit{celerite} result is only constrained well in one light curve.	
	Our results are close to the period of $\sim280$ days claimed by \citet{2016ApJ...820...20S}.
	The global significance of the period in our analysis is $> 95\%$ in CARMA model.
	It is $> 99\%$ given by \textit{celerite} model with the 30-day binning light curve, 
	while a low global significance is obtained in \textit{celerite} modeling the 14-day binning light curve.
	The corresponding local significance is higher. In particular, the local significance is $>99$\% given by both CARMA and \textit{celerite} models with the 30-day binning light curve.
    However, \citet{2019MNRAS.482.1270C} and \cite{2020A&A...634A.120A} did not find QPO signal for this source.
	For PKS 0537$-$441, the slope of the PSD we obtained is consistent with that given by \citet{2019MNRAS.482.1270C}.
	We note that a peak also appears at $\sim280$ days in the PSD of PKS 0537$-$441 given by \citet{2019MNRAS.482.1270C}, although the signal is not significant.
	The estimate for the significance of such a QPO signal should be investigated more deeply .

\begin{figure*}
	\gridline{\fig{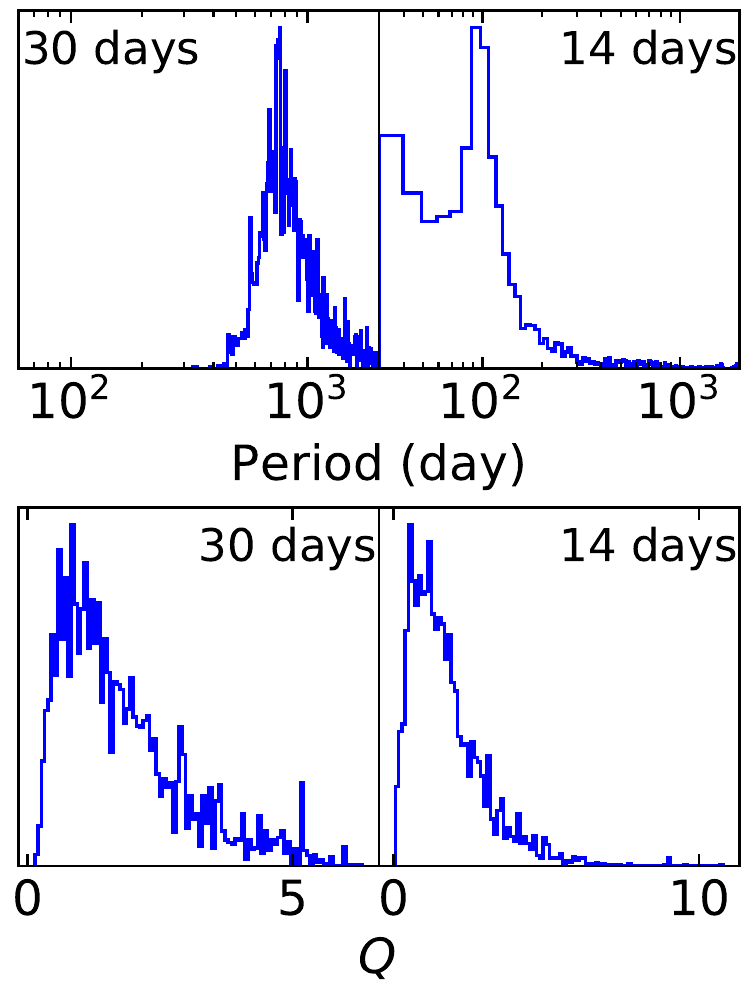}{0.38\textwidth}{(a)}
		\fig{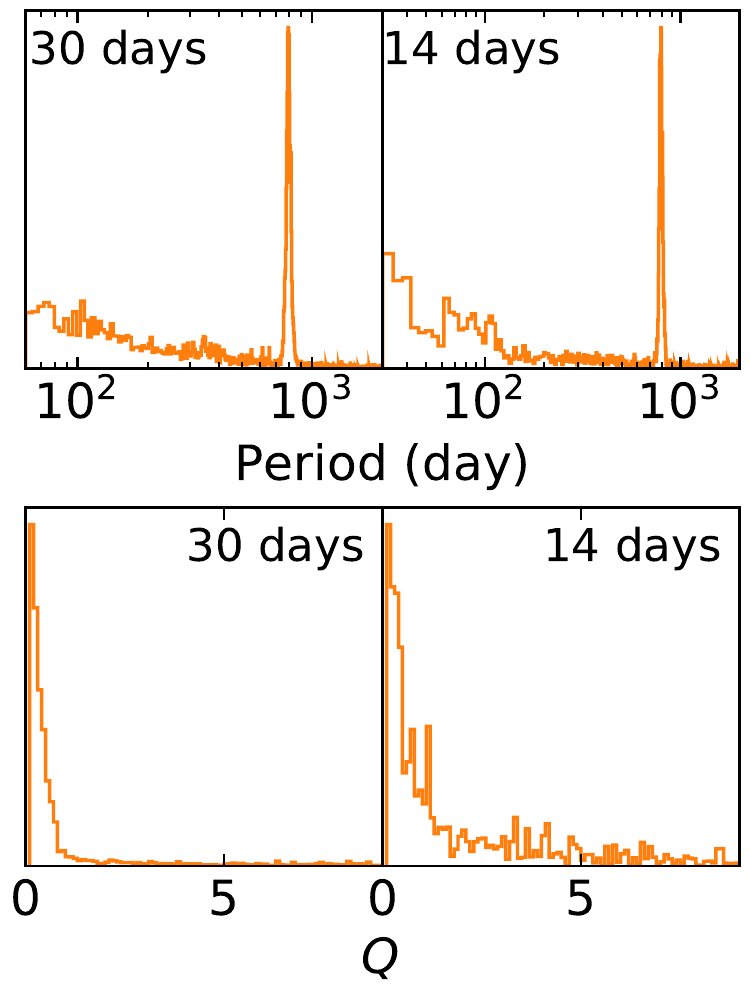}{0.38\textwidth}{(b)}}
	\caption{Marginalized posterior probability densities of period and the corresponding quality factor $Q$ for PG 1553+113. The symbols and lines are the same as those in Figure \ref{fig:pks0537post}.  \label{fig:pg1553post}}
\end{figure*}

	In the modeling results of PG 1553$+$113, possible evidence for the period of $\sim800$ days appear, with local significance of $\gtrsim95$\%. However, the global significance is 50\%-90\%.
	In the analyses of \citet{2019MNRAS.482.1270C} and \cite{2020A&A...634A.120A}, 
	they did not find significant $\gamma$-ray QPO for PG 1553$+$113.
	While \citet{2020ApJ...895..122C} claimed to confirmed the period of the PG 1553$+$113 reported by \citet{2015ApJ...813L..41A} with their Gaussian modeling method.
	The Gaussian model used in \citet{2020ApJ...895..122C}  is similar to \textit{celerite} model.
	For PG 1553$+$113, a peak at $\sim800$ days can also be found in the posterior probability density functions given by \textit{celerite} model (Figure~\ref{fig:pg1553post}), 
	but its global significance is considerably weakened by the distribution below 100 days.
	It seems that a broader posterior probability density distribution of period considered here causes different results from \citet{2020ApJ...895..122C}.

	 Here, it is concluded that possible evidence for the $\gamma$-ray QPOs in PKS 0537$-$441 and PG 1553+113 are found in our Gaussian process analyses.

	The $\gamma$-ray QPO of blazar may be caused by a periodic modulation in Doppler factor \citep{2018ApJ...867...53Y} which can be the result of jet wobbling motion or jet procession. 
	Jet wobbling motion is found for PG 1553$+$113 in the recent study of the radio emission from the jet on parsec scales \citep{2020A&A...634A..87L}. 
	However, there is no QPO found in its radio emission, suggesting that the connection between jet motion and its emission is rather complex. 
	It is hard to determine the origin of the $\gamma$-ray QPO.
	
	The $\gamma$-ray QPO in PKS 0537$-$441 appears in the early years of the \textit{Fermi} monitoring. 
	 There is an outburst along with the oscillation (Figure \ref{fig:fitlc}). 
	 \citet{2018NatCo...9.4599Z} reported the detection of a $\gamma$-ray QPO after a $\gamma$-ray outburst in the blazar PKS 2247$-$13, and the QPO vanished after about six period cycles.
	 \citet{2018NatCo...9.4599Z} explained the QPO with periodic changes of Doppler factor.
	 This scenario is also applicable to the origin of the QPO in PKS 0537$-$441.
	 The method proposed by \citet{2018ApJ...867...53Y}  can be used to test this scenario, which will be carried out in a future work.
	 On the other hand, the intrinsic origins for such a $\gamma$-ray QPO cannot be excluded, for instance a periodic change in the particle acceleration rate in a binary SMBH system \citep[e.g.,][]{2018ApJ...854...11T}.
	
	Finally, we would like to stress that \textit{celerite} models are used to successfully characterize the $\gamma$-ray light curves of blazars, 
	suggesting that  \textit{celerite} have strong potential to study AGN variability. 
	Some issues are also found in the  \textit{celerite}  modeling results, 
	for example the cause for the peaks at 40-50 days in the posterior distribution of period and the unconstrained $Q$, which likely cause the much larger uncertainties on the periods.
	A deep investigation is needed to resolve these issues, and to test numerical stability of \textit{celerite}.

	\acknowledgments
	We thank the anonymous reviewer for constructive suggestions and Yan-Rong Li (IHEP) for valuable discussions.
	We acknowledge financial support from National Key R\&D Program of China under grant No. 2018YFA0404204, 
	the National Science Foundation of China (U1738124, 11803081, U1531131 and U1738211), and the Science Foundation of Yunnan Province (NO. 2018FA004).
	D. H. Yan acknowledges financial supports from the joint foundation of Department of Science and Technology of Yunnan Province and Yunnan University [2018FY001(-003)].
	The work of D. H. Yan is also supported by the CAS Youth Innovation Promotion Association and Basic research Program of Yunnan Province (202001AW070013).
	
	\vspace{5mm}
	\facilities{Fermi (LAT)}
	
	
	\software{Fermitools-conda, carma\_pack \citep{2014ApJ...788...33K}}, celerite \citep{2017AJ....154..220F}, emcee \citep{2013PASP..125..306F}, NumPy, Matplotlib \citep{2007CSE.....9...90H}.
	
	\clearpage
	
	\bibliography{femiqpo}{}
	\bibliographystyle{aasjournal}

\startlongtable
\begin{deluxetable*}{lccccccc}
	
	\tablecaption{Modeling results for 27 blazars. The results of the 3 faint sources are listed at the end of the table. (1) source name, (2) CARMA/$celerite$ method, (3,6) the best CARMA/$celerite$ model we selected for the modeling, 
	(4,7) KS test $p$-value on the normality of the standardized residuals for the fit, (5,8) slope of the resulting PSD, and the uncertainties represent $1\sigma$ confidence intervals. \label{tab:modeling_res}}
	\tablehead{
		\colhead{Source} & \colhead{Method} & \multicolumn3c{30 days} & \multicolumn3c{14 days} \\
		\cline{3-5}
		\cline{6-8}
		&&\colhead{model} &\colhead{$p$-value} &\colhead{slope} &\colhead{model} &\colhead{$p$-value} &\colhead{slope}
	}
	\colnumbers	
	\startdata
TXS 0059+581 	&	 CARMA 	&	 $(7,6)$	&	0.459	&	 $ -1.02_{-0.33}^{+0.31}$ 	&	 $(6,5)$	&	0.080	&	 $-1.03_{-0.33}^{+0.30}$ 	\\
&	 \textit{celerite} 	&	 S$\times3$	&	0.044	&	 $ -1.36_{-0.27}^{+0.25}$ 	&	 S$\times3$	&	0.137	&	 $ -1.27_{-0.30}^{+0.21}$ 	\\
PKS 0208-512 	&	 CARMA 	&	 $(3,0)$	&	0.020	&	 $ -1.45_{-0.22}^{+0.25}$ 	&	 $(7,0)$	&	0.524	&	 $ -1.32_{-0.19}^{+0.17}$ 	\\
&	 \textit{celerite} 	&	 S$\times2$	&	0.003	&	 $ -1.33_{-0.20}^{+0.18}$ 	&	 S$\times4$	&	0.209	&	 $ -1.12_{-0.22}^{+0.19}$ 	\\
MG1 J021114+1051 	&	 CARMA 	&	 $(7,6)$	&	0.023	&	 $ -0.85_{-0.37}^{+0.35}$ 	&	 $(3,2)$	&	0.277	&	 $ -0.76_{-0.41}^{+0.38}$ 	\\
&	 \textit{celerite} 	&	 D+S	&	0.090	&	 $ -0.84_{-0.21}^{+0.17}$ 	&	 S$\times2$	&	0.085	&	 $ -0.63_{-0.21}^{+0.16}$ 	\\
PKS 0250-225 	&	 CARMA 	&	 $(1,0)$	&	0.011	&	 $ -0.72_{-0.17}^{+0.13}$ 	&	 $(1,0)$	&	0.072	&	 $ -0.83_{-0.16}^{+0.13}$ 	\\
&	 \textit{celerite} 	&	 D+S 	&	0.024	&	 $ -0.72\pm0.16$ 	&	 S 	&	0.002	&	 $ -0.71_{-0.14}^{+0.12}$ 	\\
PKS 0301-243 	&	 CARMA 	&	 $(4,2)$	&	0.308	&	 $ -0.65_{-0.30}^{+0.48}$ 	&	 $(3,1)$	&	0.379	&	 $ -0.76_{-0.45}^{+0.37}$ 	\\
&	 \textit{celerite} 	&	 D+S	&	0.000	&	 $ -0.49_{-0.32}^{+0.15}$ 	&	 D+S$\times2$	&	0.000	&	 $ -0.58_{-0.16}^{+0.12}$ 	\\
PKS 0426-380 	&	 CARMA 	&	 $(7,5)$	&	0.013	&	 $ -1.16_{-0.22}^{+0.21}$ 	&	 $(6,4)$	&	0.181	&	 $ -1.20_{-0.19}^{+0.18}$ 	\\
&	 \textit{celerite} 	&	 D+S 	&	0.008	&	 $ -1.29_{-0.16}^{+0.14}$ 	&	 S$\times2$	&	0.001	&	 $ -1.12_{-0.17}^{+0.14}$ 	\\
PKS 0447-439 	&	 CARMA 	&	 $(2,1)$	&	0.461	&	 $ -1.04_{-0.29}^{+0.23}$ 	&	 $(3,1)$	&	0.808	&	 $ -1.02_{-0.26}^{+0.22}$ 	\\
&	 \textit{celerite} 	&	 S	&	0.260	&	 $ -1.00_{-0.25}^{+0.21}$ 	&	 D+S	&	0.431	&	 $ -0.95_{-0.22}^{+0.17}$ 	\\
PKS 0454-234 	&	 CARMA 	&	 $(3,1)$	&	0.090	&	 $-1.27_{-0.29}^{+0.24}$ 	&	 $(4,2)$	&	0.077	&	 $ -1.21_{-0.20}^{+0.15}$ 	\\
&	 \textit{celerite} 	&	 D+S	&	0.139	&	 $ -1.35_{-0.17}^{+0.15}$ 	&	 D+S$\times2$	&	0.006	&	 $ -1.04_{-0.13}^{+0.10}$ 	\\
S3 0458-02 	&	 CARMA 	&	 $(1,0)$	&	0.001	&	 $ -0.69_{-0.18}^{+0.13}$ 	&	 $(1,0)$	&	0.001	&	 $ -0.55_{-0.10}^{+0.09}$ 	\\
&	 \textit{celerite} 	&	 D+S	&	0.001	&	 $ -0.71_{-0.17}^{+0.16}$ 	&	 S	&	0.000	&	 $ -0.50_{-0.11}^{+0.09}$ 	\\
TXS 0518+211 	&	 CARMA 	&	 $(3,1)$	&	0.455	&	 $ -1.31_{-0.28}^{+0.25}$ 	&	 $(2,1)$	&	0.599	&	 $ -1.22_{-0.32}^{+0.26}$ 	\\
&	 \textit{celerite} 	&	 S 	&	0.671	&	 $ -1.23_{-0.28}^{+0.22}$ 	&	 S	&	0.422	&	 $ -1.16_{-0.27}^{+0.21}$ 	\\
PKS 0537-441 	&	 CARMA 	&	 $(5,1)$	&	0.100	&	 $ -1.67_{-0.24}^{+0.20}$ 	&	 $(5,2)$	&	0.224	&	 $ -1.72_{-0.27}^{+0.23}$ 	\\
&	 \textit{celerite} 	&	 D+S	&	0.173	&	 $ -1.67_{-0.12}^{+0.14}$ 	&	 D+S$\times2$	&	0.097	&	 $ -1.70_{-0.15}^{+0.14}$ 	\\
S5 0716+714 	&	 CARMA 	&	 $(3,0)$	&	0.133	&	 $ -0.90_{-0.22}^{+0.21}$ 	&	 $(5,0)$	&	0.002	&	 $ -0.85_{-0.23}^{+0.21}$ 	\\
&	 \textit{celerite} 	&	 S	&	0.183	&	 $ -0.94_{-0.13}^{+0.12}$ 	&	 S 	&	0.000	&	 $ -0.90\pm0.21$ 	\\
PKS 0805-077 	&	 CARMA 	&	 $(5,4)$	&	0.119	&	 $ -0.91_{-0.25}^{+0.23}$ 	&	 $(4,1)$	&	0.054	&	 $ -1.19_{-0.18}^{+0.16}$ 	\\
&	 \textit{celerite} 	&	 D+S 	&	0.034	&	 $ -0.99_{-0.22}^{+0.18}$ 	&	 D+S$\times2$ 	&	0.005	&	 $ -0.55_{-0.14}^{+0.12}$ 	\\
OJ 014 	&	 CARMA 	&	 $(2,1)$	&	0.972	&	 $ -1.23_{-0.41}^{+0.45}$ 	&	 $(2,1)$	&	0.539	&	 $ -0.47_{-0.55}^{+0.47}$ 	\\
&	 \textit{celerite} 	&	 S	&	0.794	&	 $ -1.35_{-0.50}^{+0.48}$ 	&	 S	&	0.491	&	 $ -0.64_{-0.58}^{+0.44}$ 	\\
S4 0814+42 	&	 CARMA 	&	 $(2,1)$	&	0.686	&	 $ -1.48_{-0.26}^{+0.33}$ 	&	 $(2,0)$	&	0.702	&	 $ -1.50_{-0.24}^{+0.39}$ 	\\
&	 \textit{celerite} 	&	 S	&	0.782	&	 $ -1.48_{-0.42}^{+0.38}$ 	&	 D+S$\times2$	&	0.754	&	 $ -1.64_{-0.32}^{+0.43}$ 	\\
4C +01.28 	&	 CARMA 	&	 $(7,6)$	&	0.819	&	 $-0.81_{-0.28}^{+0.22}$ 	&	 $(7,6)$	&	0.515	&	 $-0.82_{-0.28}^{+0.24}$ 	\\
&	 \textit{celerite} 	&	 D+S	&	0.230	&	 $-0.80_{-0.21}^{+0.17}$	&	 D+S	&	0.049	&	 $-0.68_{-0.20}^{+0.15}$	\\
S4 1144+40 	&	 CARMA 	&	 $(6,5)$	&	0.179	&	 $ -1.34_{-0.21}^{+0.19}$ 	&	 $(5,2)$	&	0.564	&	 $ -1.17\pm0.22$ 	\\
&	 \textit{celerite} 	&	 D+S 	&	0.437	&	 $ -1.15_{-0.26}^{+0.18}$ 	&	  D+S 	&	0.030	&	 $ -0.90_{-0.23}^{+0.16}$ 	\\
PG 1246+586 	&	 CARMA 	&	 $(4,1)$	&	0.569	&	 $ -0.68_{-0.70}^{+0.69}$ 	&	 $(7,6)$	&	0.325	&	 $ -0.45_{-0.68}^{+0.56}$ 	\\
&	 \textit{celerite} 	&	 S 	&	0.952	&	 $ -0.79_{-0.43}^{+0.29}$ 	&	 S$\times2$	&	0.826	&	 $ -0.88_{-0.57}^{+0.45}$ 	\\
PG 1553+113 	&	 CARMA 	&	 $(4,1)$	&	0.776	&	 $ -1.05_{-0.47}^{+0.45}$ 	&	 $(3,0)$	&	0.401	&	 $ -1.12_{-0.39}^{+0.32}$ 	\\
&	 \textit{celerite} 	&	 S$\times4$	&	0.436	&	 $ -0.70_{-0.57}^{+0.41}$ 	&	 S$\times4$	&	0.565	&	 $ -0.72_{-0.45}^{+0.49}$ 	\\
TXS 1902+556 	&	 CARMA 	&	 $(2,1)$	&	0.814	&	 $ -0.78_{-0.56}^{+0.44}$ 	&	 $(6,4)$	&	0.086	&	 $ -0.41_{-0.63}^{+0.35}$ 	\\
&	 \textit{celerite} 	&	 S$\times2$	&	0.952	&	 $ -0.81_{-0.66}^{+0.43}$ 	&	 S$\times2$	&	0.300	&	 $ -0.84_{-0.75}^{+0.47}$ 	\\
PKS 2052-477 	&	 CARMA 	&	 $(5,0)$	&	0.021	&	 $ -1.15_{-0.26}^{-0.32}$ 	&	 $(6,4)$	&	0.075	&	 $ -1.02_{-0.24}^{+0.21}$ 	\\
&	 \textit{celerite} 	&	 D+S$\times2$	&	0.090	&	 $ -1.09_{-0.38}^{+0.25}$ 	&	 S$\times4$	&	0.249	&	 $ -0.91_{-0.23}^{+0.14}$ 	\\
PKS 2155-304 	&	 CARMA 	&	 $(1,0)$	&	0.050	&	 $ -1.00_{-0.24}^{+0.18}$ 	&	 $(5,0)$	&	0.119	&	 $ -0.95_{-0.25}^{+0.21}$ 	\\
&	 \textit{celerite} 	&	 S	&	0.026	&	 $ -0.81_{-0.17}^{+0.13}$ 	&	 S$\times2$	&	0.265	&	 $ -0.79_{-0.15}^{+0.13}$ 	\\
BL Lacertae 	&	 CARMA 	&	 $(2,1)$	&	0.332	&	 $ -1.30_{-0.20}^{+0.25}$ 	&	 $(2,1)$	&	0.039	&	 $ -1.23_{-0.14}^{+0.13}$ 	\\
&	 \textit{celerite} 	&	 D+S	&	0.341	&	 $ -1.21_{-0.23}^{+0.16}$ 	&	 D+S 	&	0.005	&	 $ -1.20_{-0.21}^{+0.18}$	\\
PKS 2255-282 	&	 CARMA 	&	 $(2,1)$	&	0.165	&	 $ -1.43\pm0.22$ 	&	 $(7,4)$	&	0.558	&	 $ -1.21_{-0.29}^{+0.28}$ 	\\
&	 \textit{celerite} 	&	 S$\times2$	&	0.302	&	 $ -1.33_{-0.28}^{+0.23}$ 	&	 D+S	&	0.165	&	 $ -1.01_{-0.23}^{+0.19}$ 	\\
\hline
\multicolumn8c{Three faint sources} \\
\hline
GB6 J0043+3426 	&	 CARMA 	&	 $(7,5)$	&	0.446	&	 $-0.98_{-0.56}^{+0.48}$ 	&	 $(6,3)$	&	0.028	&	 $-0.77_{-0.52}^{+0.50}$ 	\\
&	 \textit{celerite} 	&	 S$\times2$	&	0.805	&	 $ -1.11_{-0.58}^{+0.46}$ 	&	 S$\times2$ 	&	0.404	&	 $ -1.10_{-0.65}^{+0.64}$ 	\\
MG2 J130304+2434 	&	 CARMA 	&	 $(1,0)$	&	0.466	&	 $ -0.77_{-0.27}^{-0.20}$ 	&	 $(3,2)$	&	0.059	&	 $ -0.49_{-0.28}^{+0.23}$ 	\\
&	 \textit{celerite} 	&	 D+S	&	0.078	&	 $ -0.61_{-0.22}^{+0.17}$ 	&	 S$\times3$	&	0.142	&	 $ -0.50_{-0.21}^{+0.15}$ 	\\
87GB 164812.2+524023 	&	 CARMA 	&	 $(2,1)$	&	0.352	&	 $ -0.68_{-0.62}^{+0.67}$ 	&	 $(7,3)$	&	0.048	&	 $ -0.30_{-0.86}^{+0.40}$ 	\\
&	 \textit{celerite} 	&	 S	&	0.921	&	 $ -0.89_{-0.72}^{+0.69}$ 	&	 D+S	&	0.373	&	 $ -1.51_{-0.45}^{+0.76}$ 	\\
	\enddata
\end{deluxetable*}
\begin{deluxetable*}{lccccccccc}
	\tabletypesize{\footnotesize}
	\tablecaption{Period, quality factor $Q$, and the corresponding significance estimated by the posterior analyses of CARMA and $celerite$ modeling for PKS 0537-441 and PG 1553+113. L. S. and G. S are the abbreviations of local significance and global significance respectively. The uncertainties represent $1\sigma$ confidence intervals. \label{tab:post}}
	\tablehead{\colhead{Source} & \colhead{Method} & \multicolumn4c{30 days} & \multicolumn4c{14 days} \\
		\cline{3-6} 
		\cline{7-10}
		&&\colhead{Period} & \colhead{$Q$} & \colhead{L. S.}&\colhead{G. S.}&\colhead{Period} & \colhead{$Q$} & \colhead{L. S.}&\colhead{G. S.}	\\
		&& 	\colhead{(days)} &&\colhead{(\%)}&\colhead{(\%)}& \colhead{(days)} &&\colhead{(\%)}&\colhead{(\%)}
	}
	\startdata
	PKS 0537-441 & CARMA & $ 233.42^{+27.21}_{-23.01}$& $ 2.77^{+1.62}_{-1.10}$ & $99.337$ & $96.685$ &$280.57 ^{+63.44}_{-40.43}$& $ 2.03^{+0.87}_{-0.79}$ & $98.453$ & $95.907 $ \\
	& \textit{celerite} & $273.10 ^{+13.32}_{-13.52}$& $9.20 ^{+25.27}_{-5.45}$ & $99.911$ & $99.603$ & $270.79 ^{+282.42}_{-198.32}$& $ 6.70^{+225.70}_{-6.70}$  & $93.608$ & $49.264$ \\
	PG 1553+113 & CARMA & $840.25^{+337.60}_{-164.38}$ & $1.46^{+1.43}_{-0.77}$ & $94.538$ & $90.772$ & $101.94^{+134.36}_{-45.39}$ & $1.42^{+1.36}_{-0.82}$  & --- & ---\\
	& \textit{celerite} & $ 791.98^{+19.47}_{-289.75}$ & $62.47 ^{+64016.75}_{-62.47}$  & $96.419$ & $48.005$ & $780.81 ^{+30.37}_{-675.87}$ & $ 299.14^{+102933.42}_{-299.14}$  & $98.161$ & $52.380$ \\
	\enddata
\end{deluxetable*}

\begin{figure*}
	
	\includegraphics[width=0.33\textwidth]{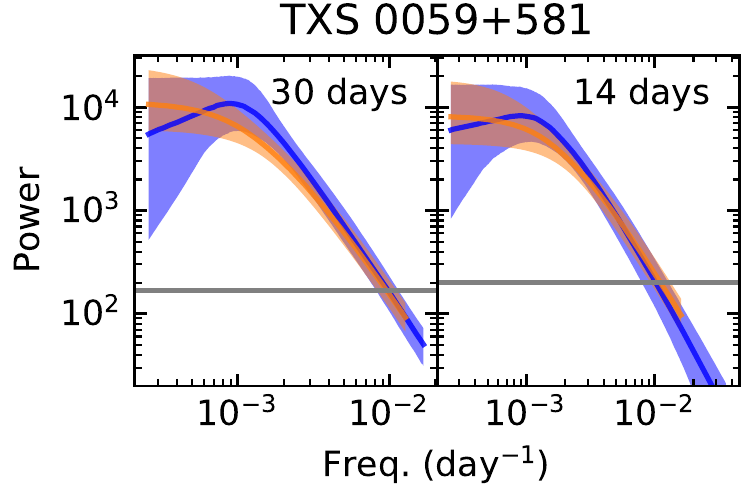}
	\includegraphics[width=0.33\textwidth]{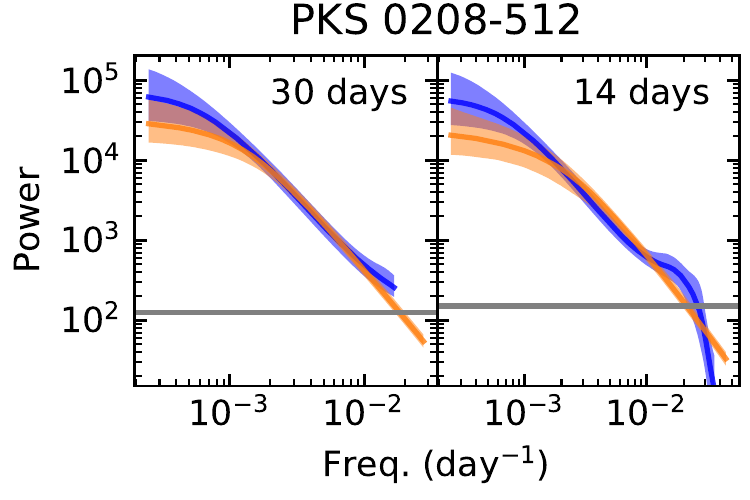}
	\includegraphics[width=0.33\textwidth]{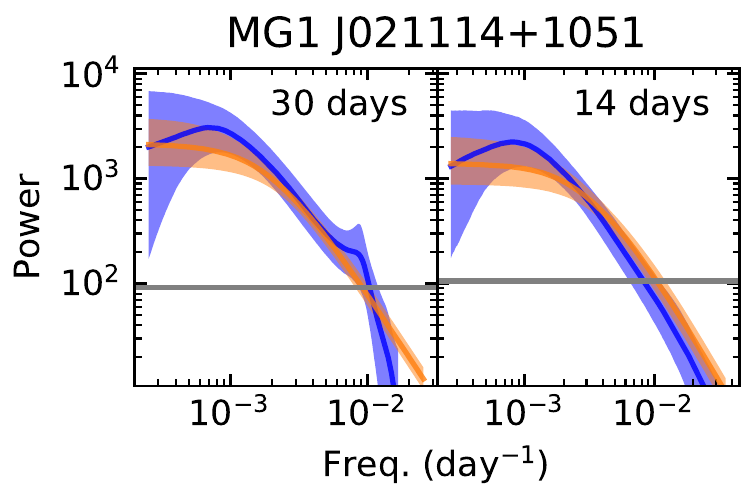}
	\includegraphics[width=0.33\textwidth]{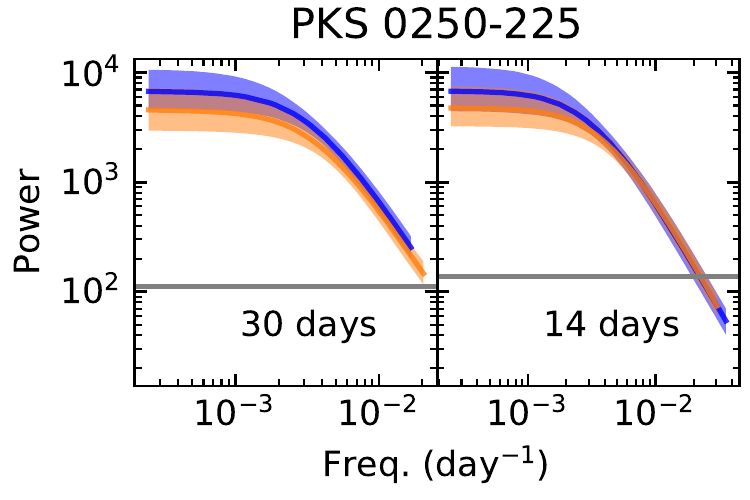}
	\includegraphics[width=0.33\textwidth]{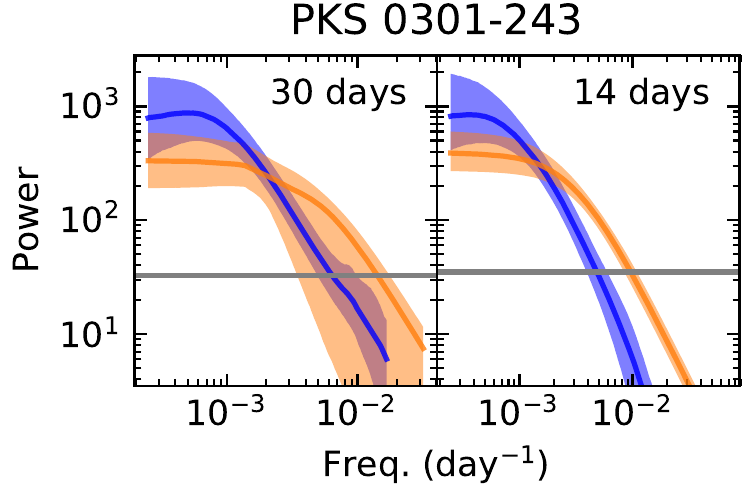}
	\includegraphics[width=0.33\textwidth]{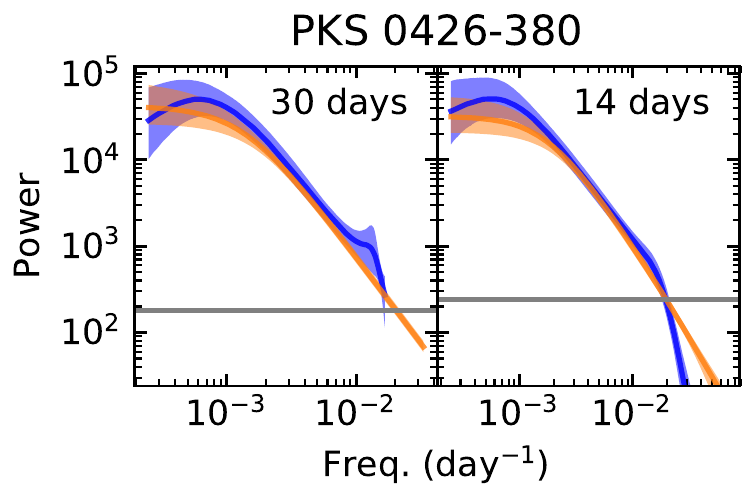}
	\includegraphics[width=0.33\textwidth]{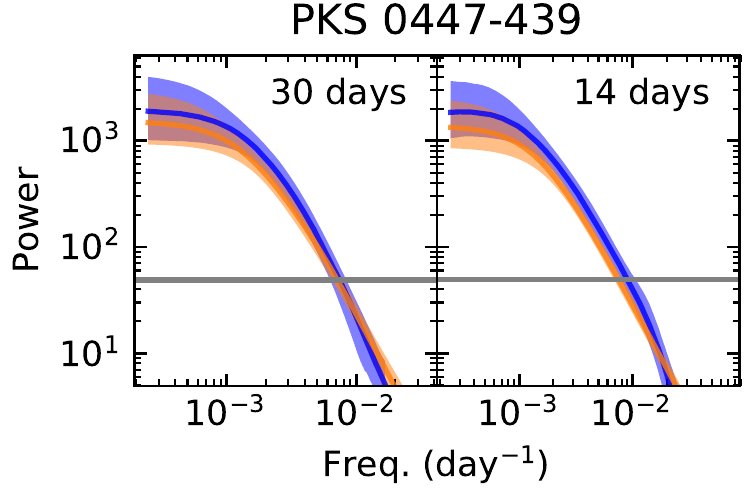}
	\includegraphics[width=0.33\textwidth]{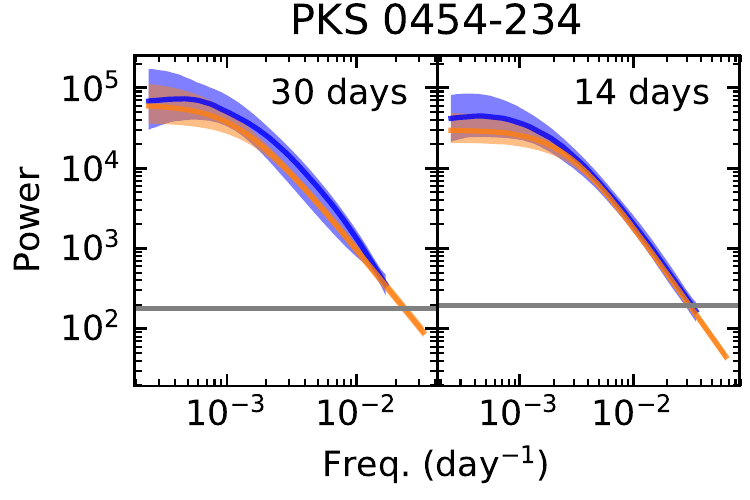}
	\includegraphics[width=0.33\textwidth]{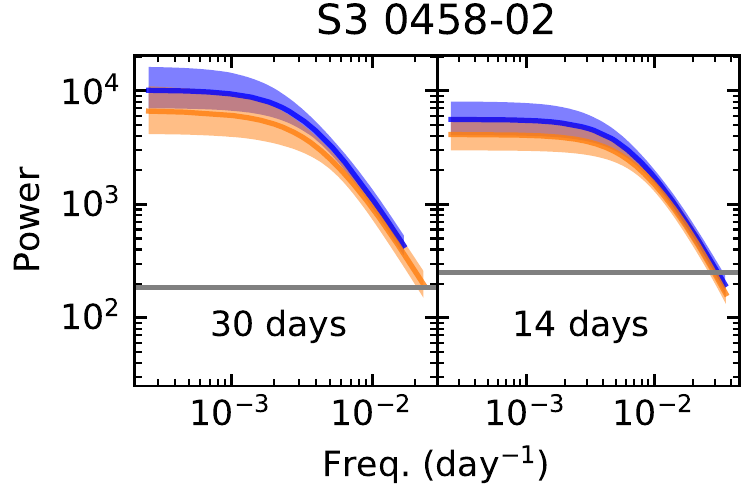}
	\includegraphics[width=0.33\textwidth]{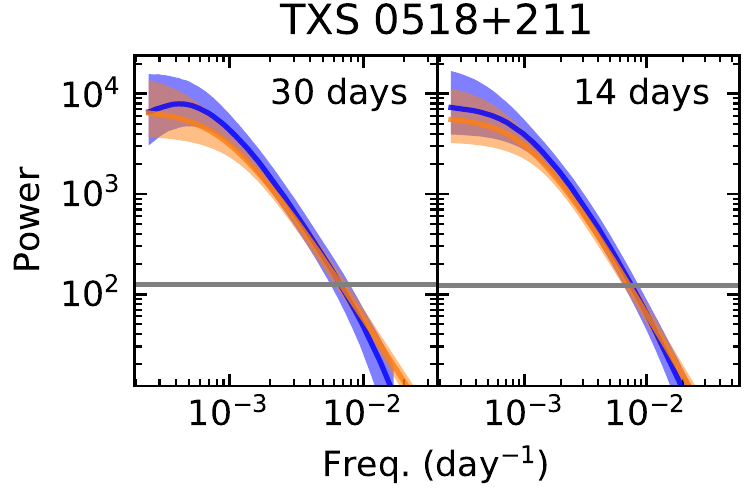}
	\includegraphics[width=0.33\textwidth]{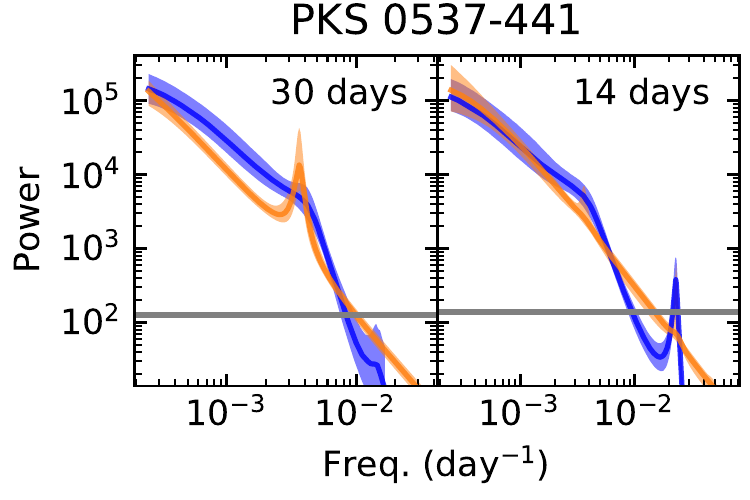}
	\includegraphics[width=0.33\textwidth]{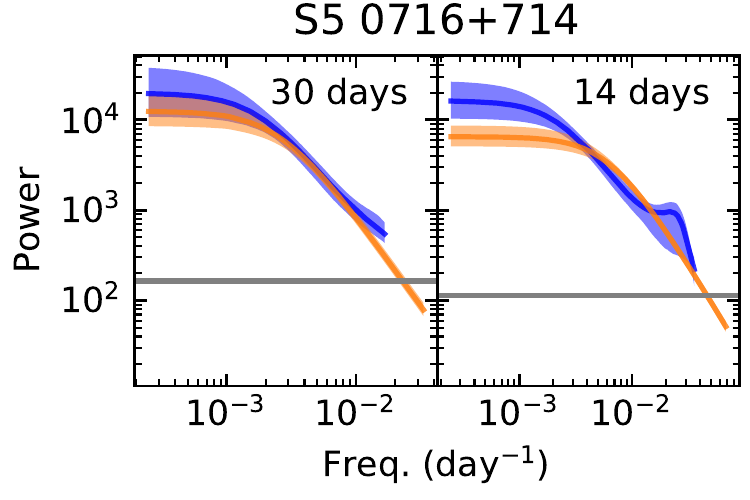}
	\includegraphics[width=0.33\textwidth]{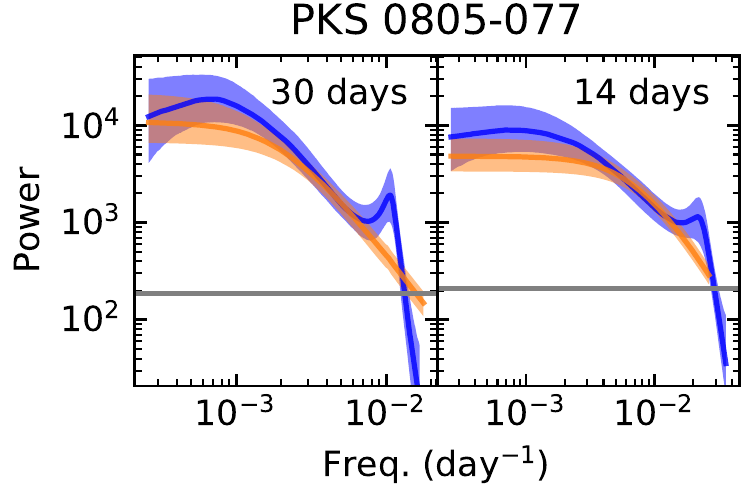}
	\includegraphics[width=0.33\textwidth]{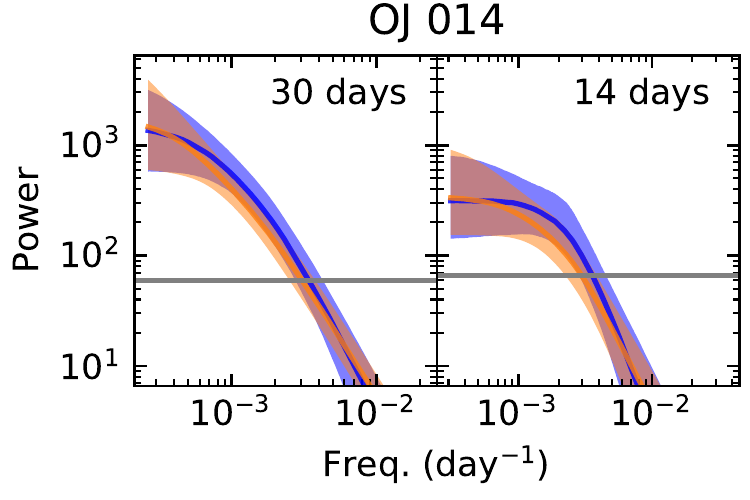}
	\includegraphics[width=0.33\textwidth]{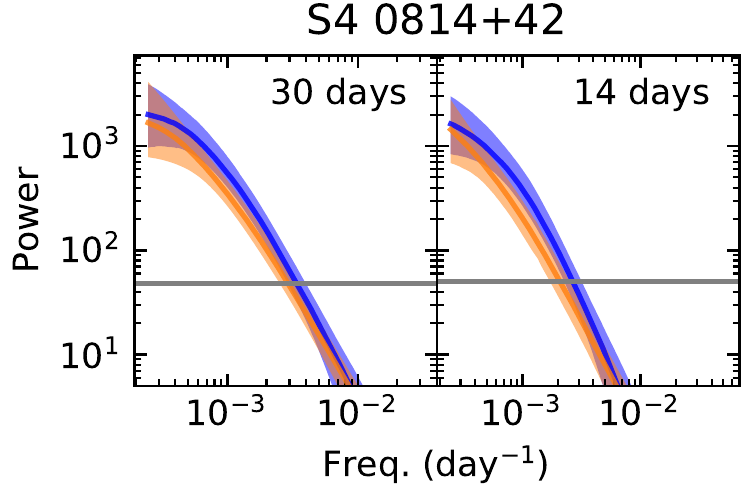}

	\caption{PSDs of $\gamma$-ray light curves of the 24 blazars. For each blazar, the left panel is the PSD for the 30-day light curve, and the right panel is the PSD for the 14-day light curve. 
	Blue and bright orange regions are the $1\sigma$ confidence intervals for CARMA and \textit{celerite} respectively. The gray horizontal lines are the measurement noise levels calculated using Equation (\ref{eq:noise}). \label{fig:allpsd}}
\end{figure*}
\begin{figure*}
	\figurenum{3}
	\includegraphics[width=0.33\textwidth]{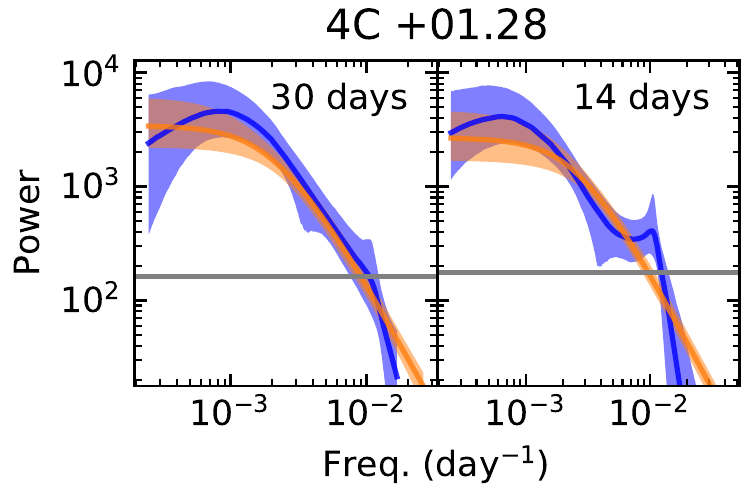}
	\includegraphics[width=0.33\textwidth]{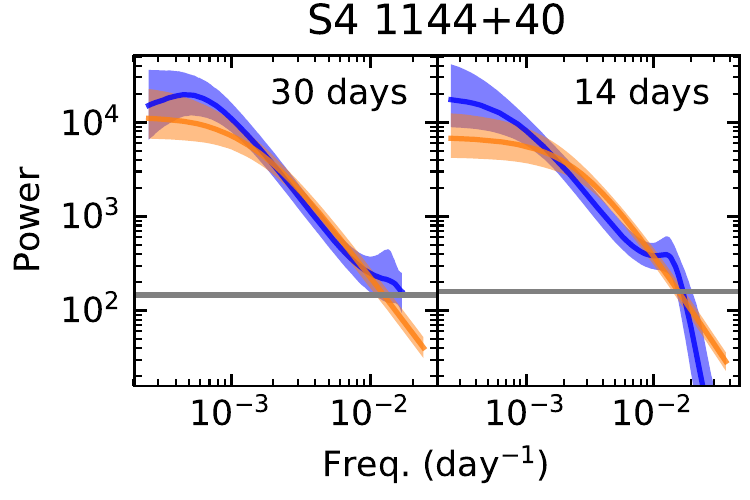}
	\includegraphics[width=0.33\textwidth]{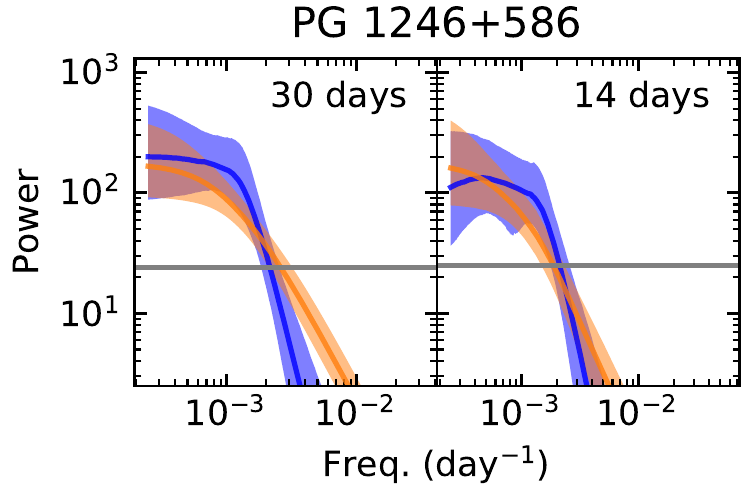}	
	\includegraphics[width=0.33\textwidth]{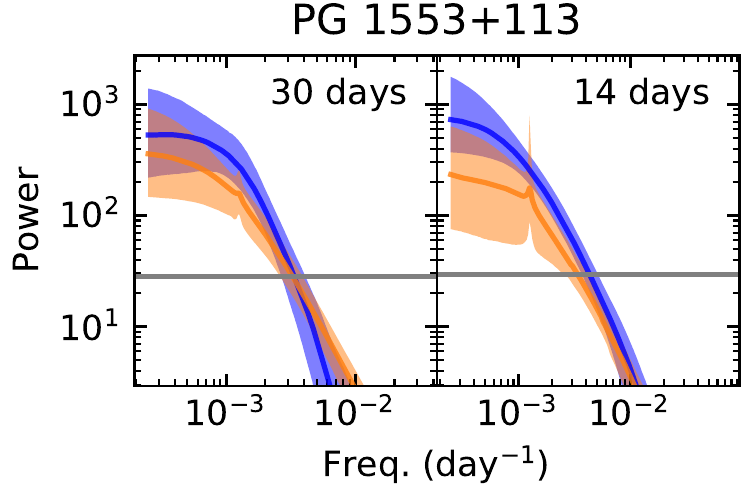}	
	\includegraphics[width=0.33\textwidth]{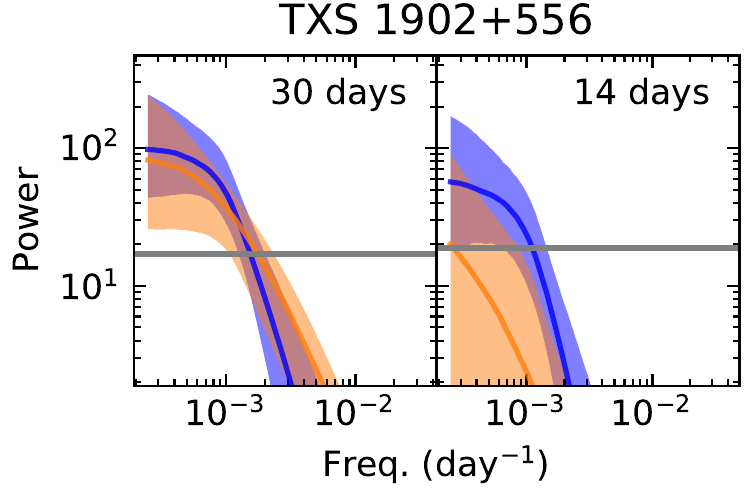}
	\includegraphics[width=0.33\textwidth]{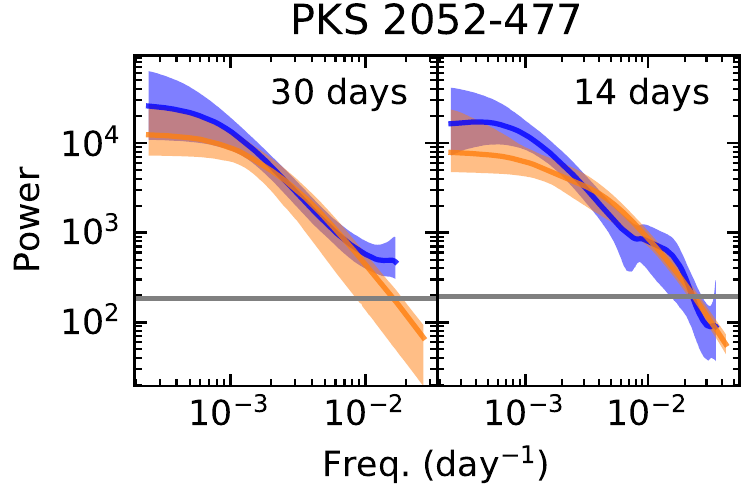}
	\includegraphics[width=0.33\textwidth]{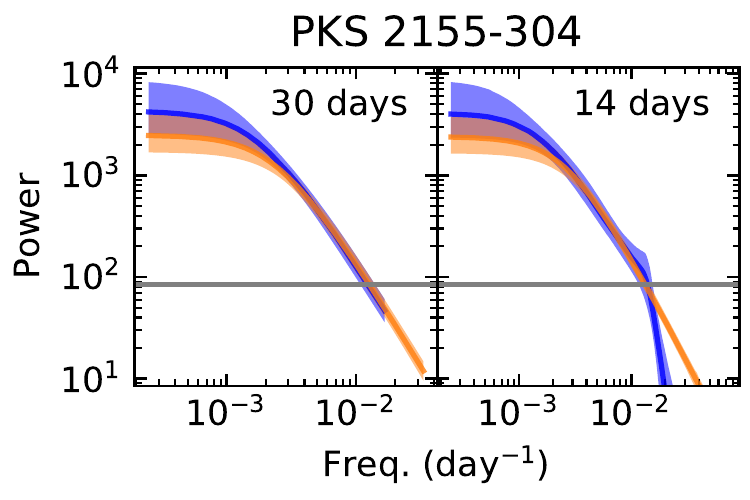}
	\includegraphics[width=0.33\textwidth]{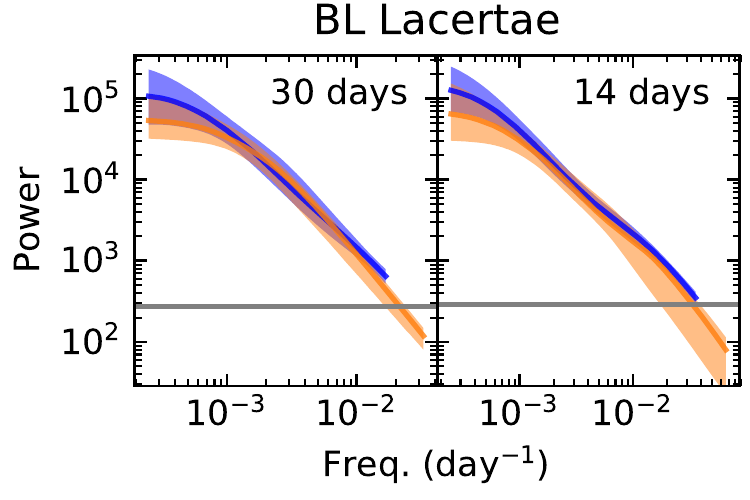}
	\includegraphics[width=0.33\textwidth]{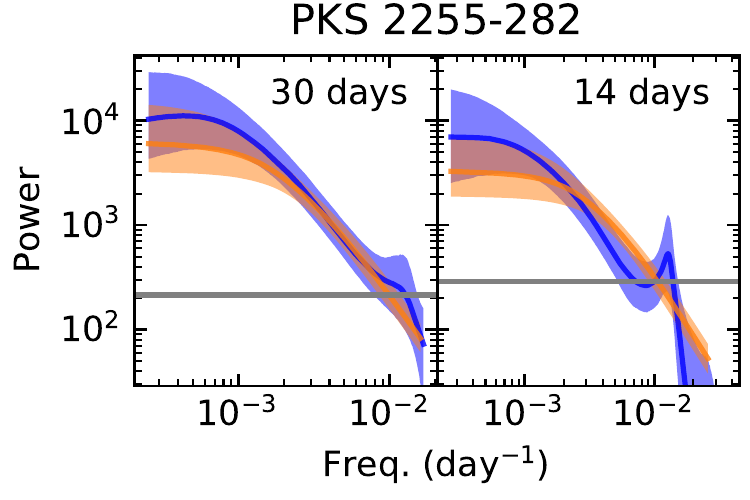}
	\caption{(\textit{continued})}
\end{figure*}

\begin{figure*}
	\centering
	\includegraphics[width=0.329\textwidth]{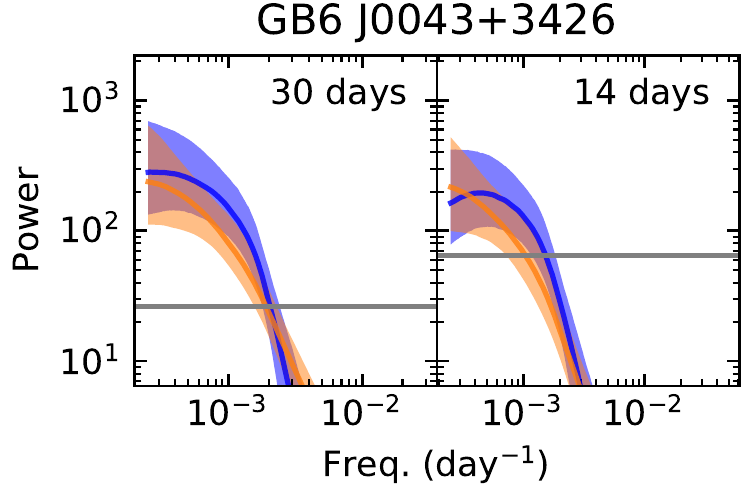}
	\includegraphics[width=0.33\textwidth]{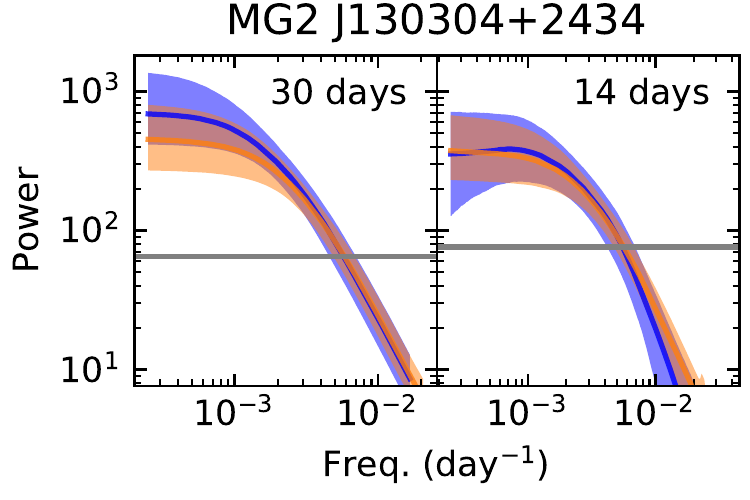}
	\includegraphics[width=0.33\textwidth]{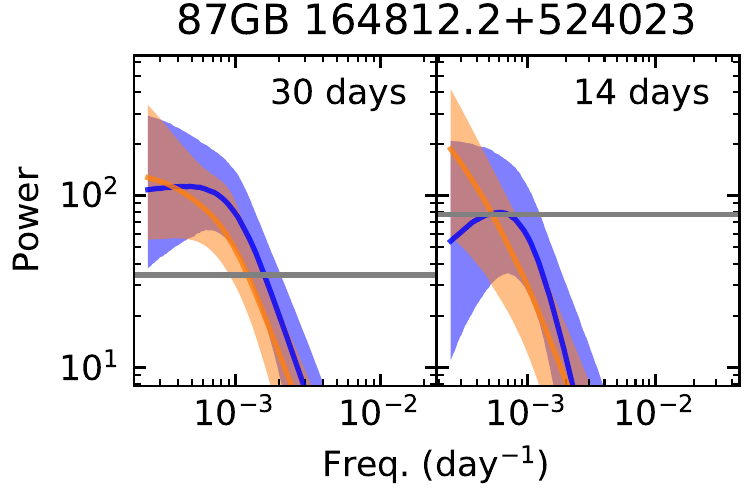}
	\caption{PSDs for the 3 faint sources. The symbols and lines are the same as those in Figure \ref{fig:allpsd}. \label{fig:psd_3}}
\end{figure*}


\clearpage
\appendix
\label{app}
\section{Analysis of the light curves with logarithmic flux} \label{append:a}

\begin{figure*}
	\includegraphics[width=0.5\textwidth]{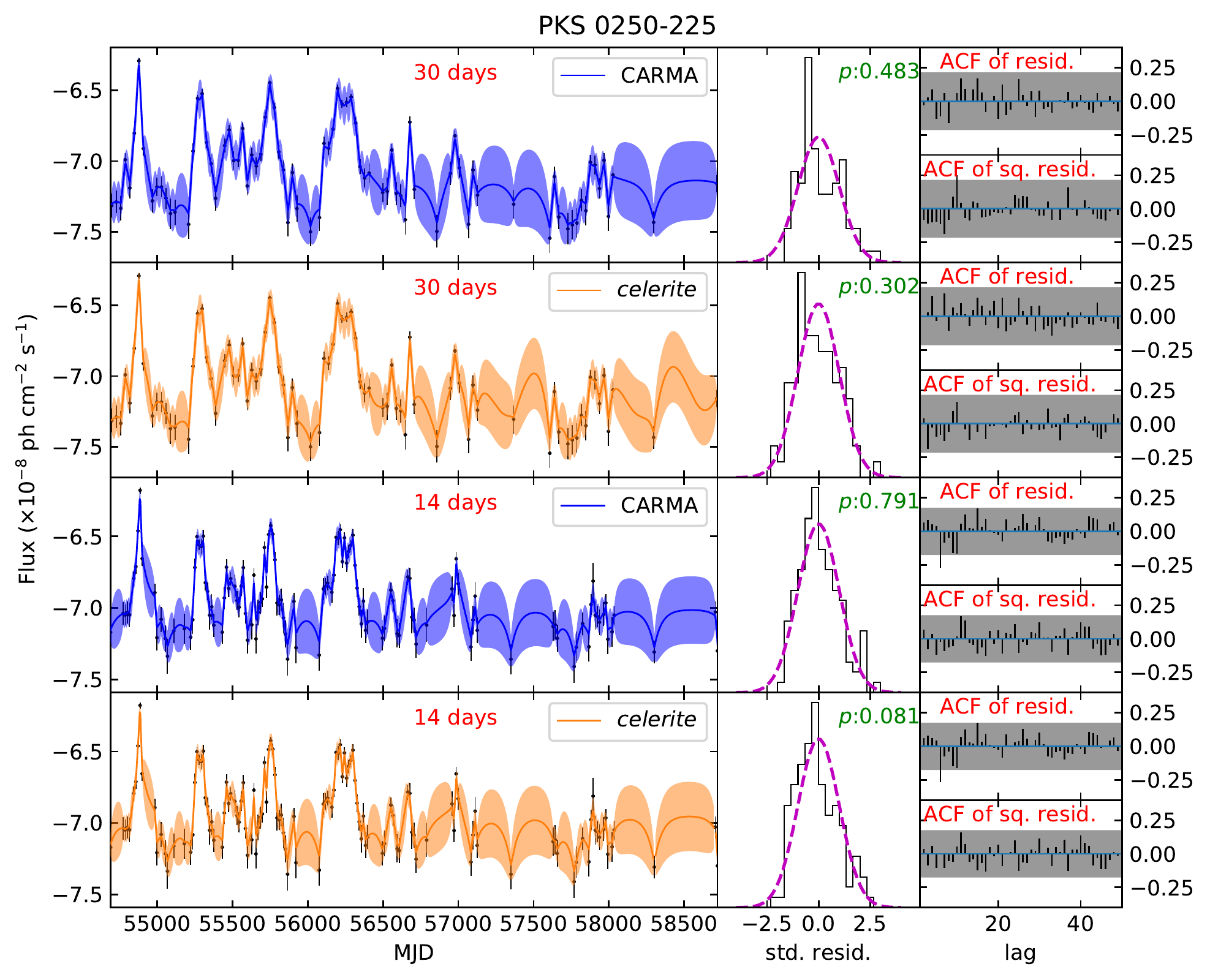}
	\includegraphics[width=0.5\textwidth]{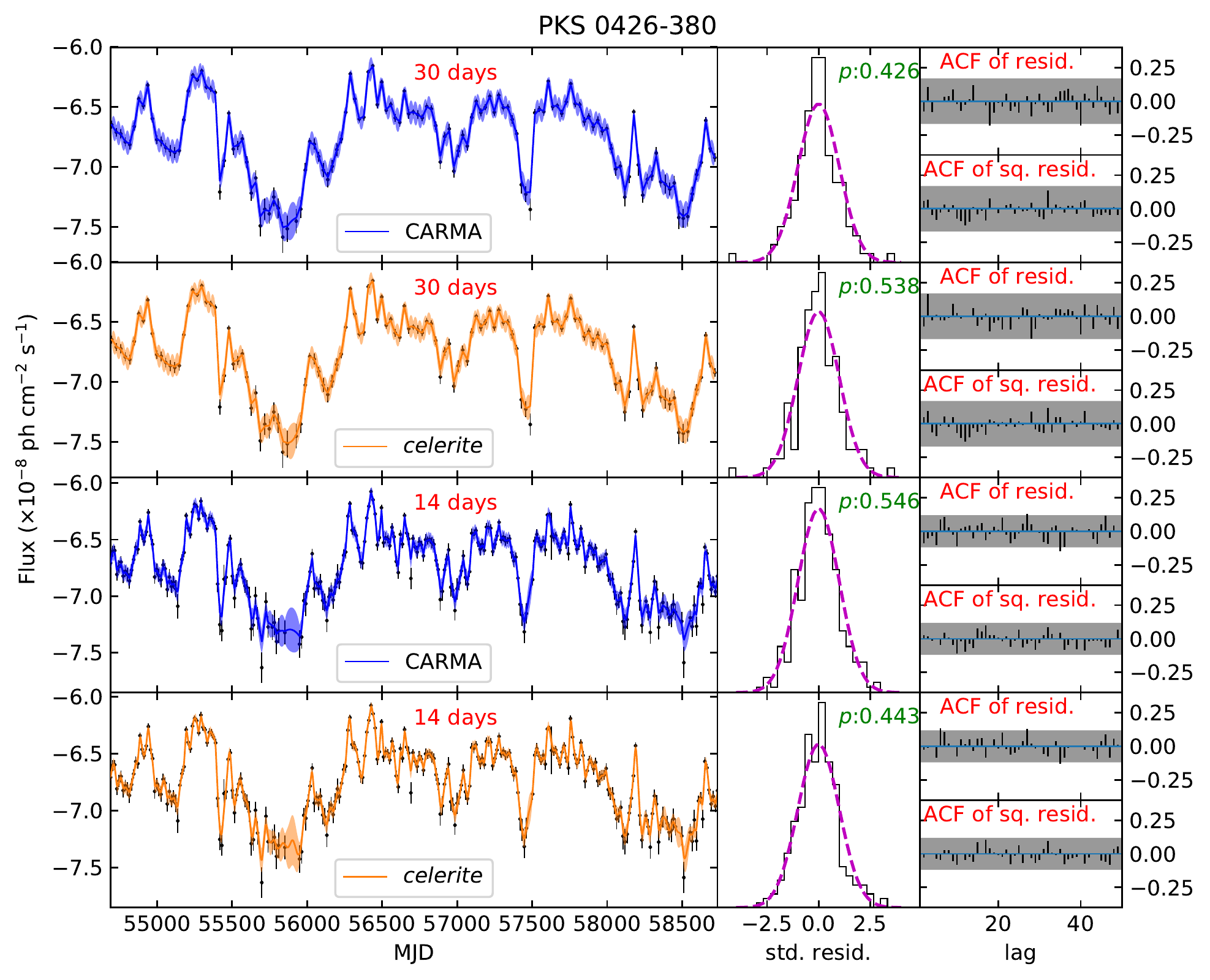}
	\includegraphics[width=0.5\textwidth]{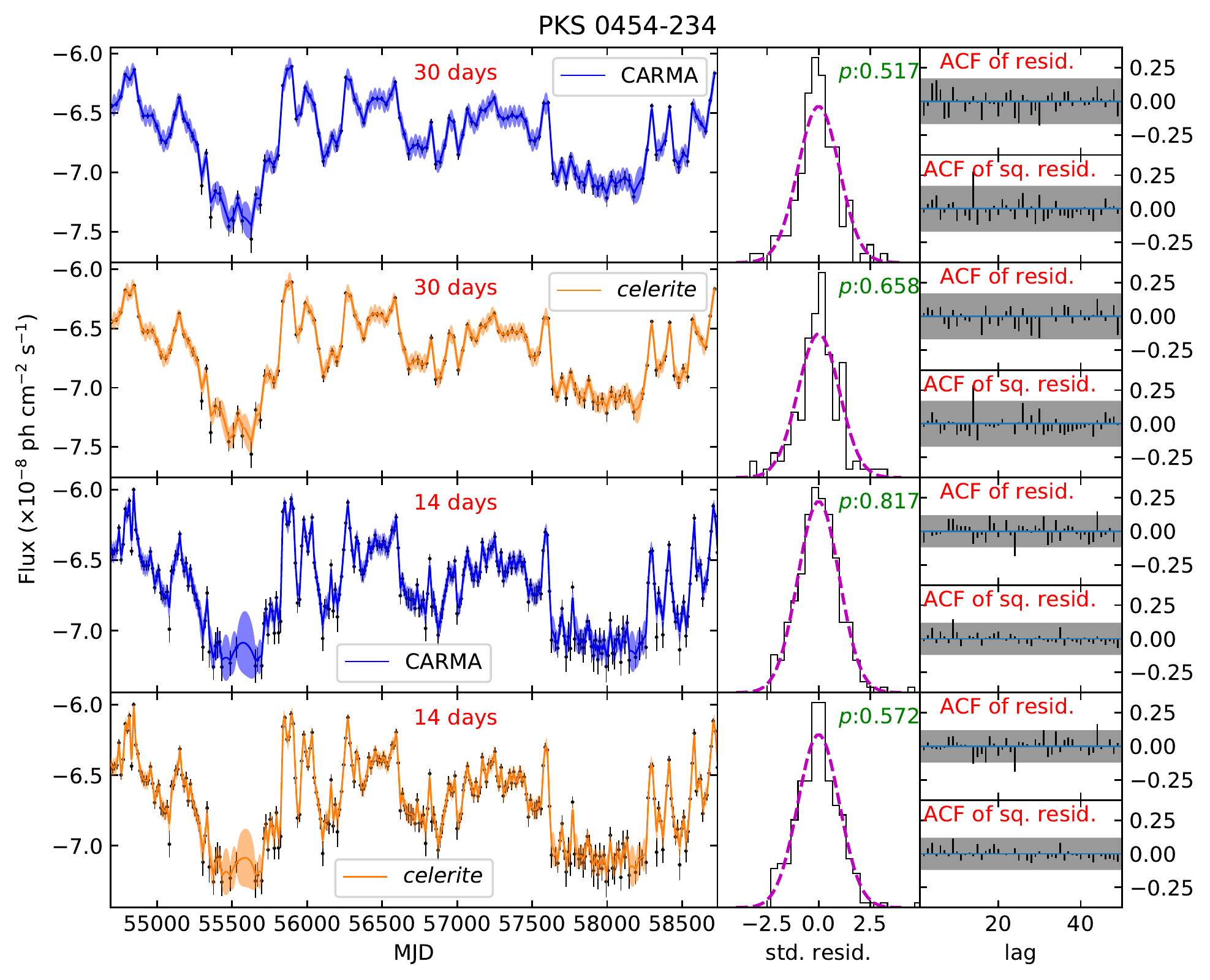}
	\includegraphics[width=0.5\textwidth]{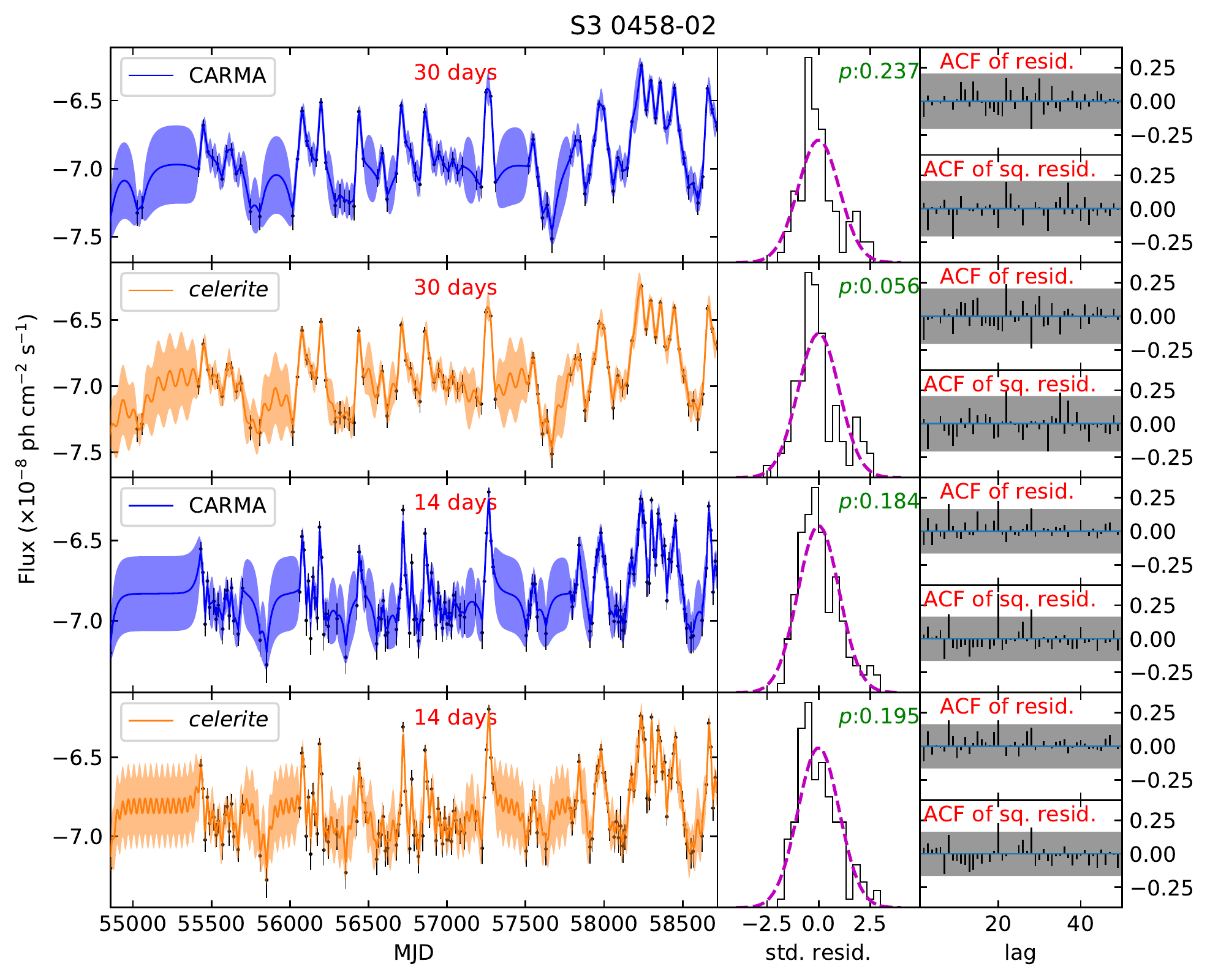}
	\includegraphics[width=0.5\textwidth]{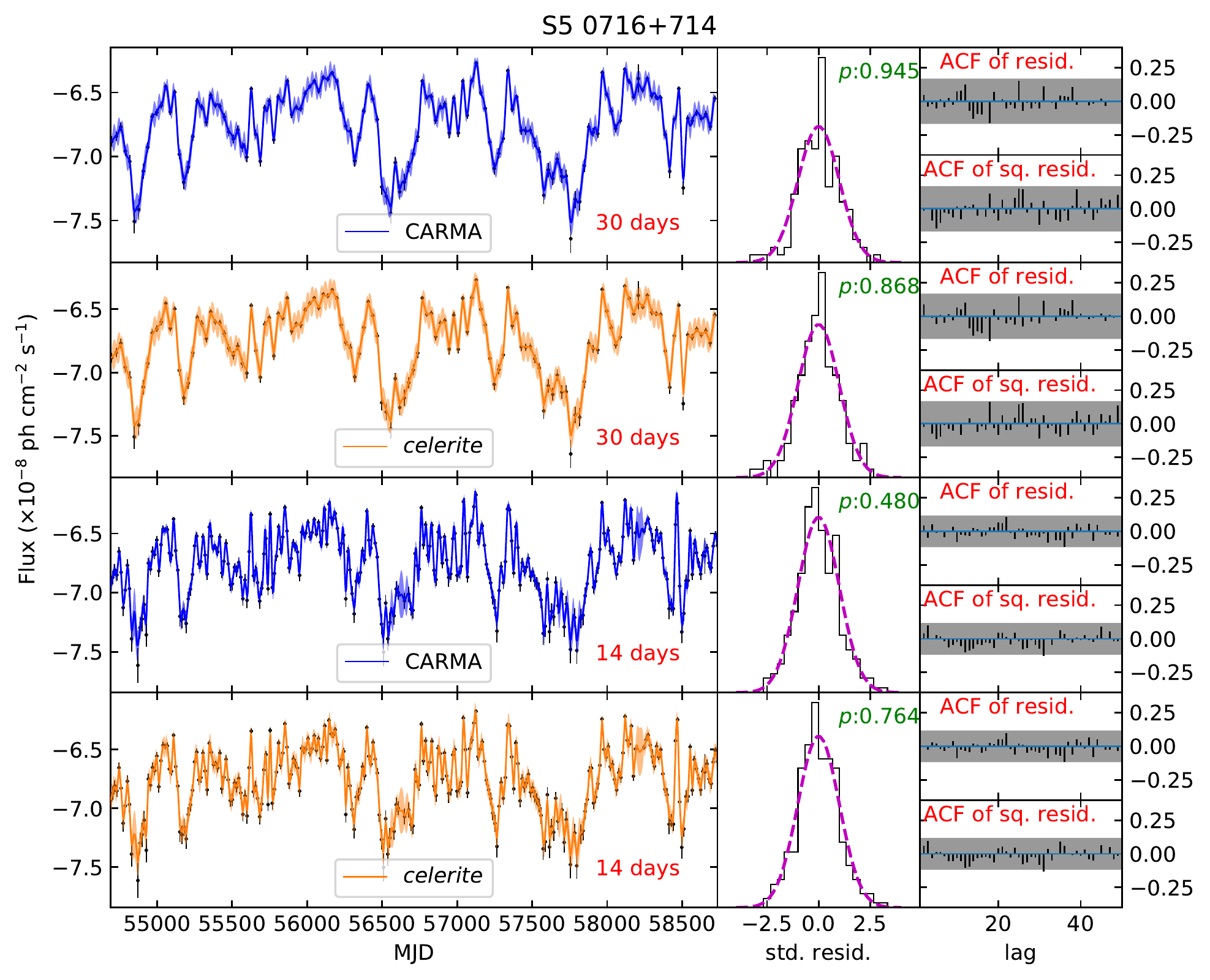}
	\includegraphics[width=0.5\textwidth]{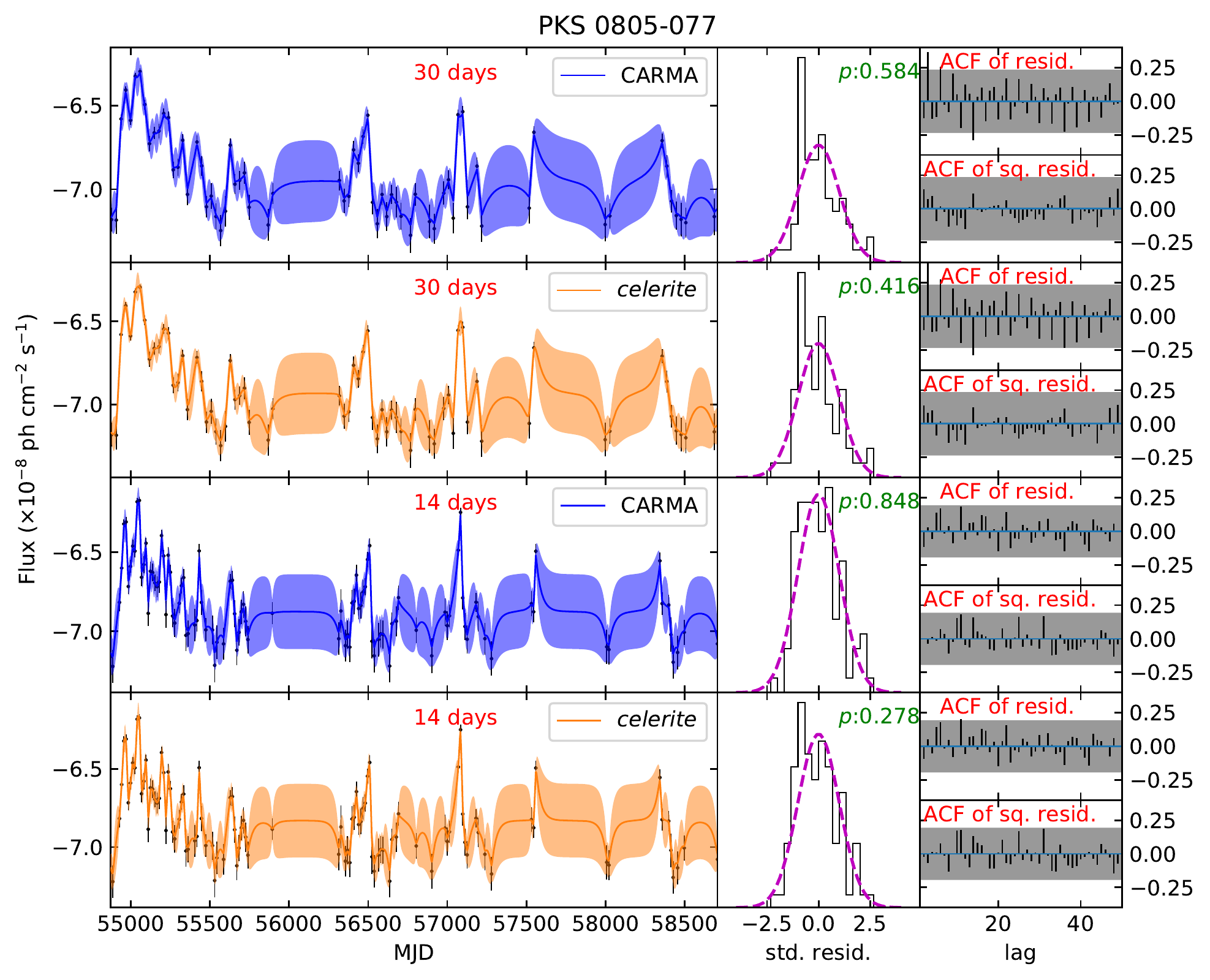}
	\caption{Fitting results for the logarithm of light curves for the 6 sources. The symbols and lines are the same as those in Figure \ref{fig:fitlc}. \label{appendfig:fitresult}}
\end{figure*}

\begin{figure*}
	\includegraphics[width=0.33\textwidth]{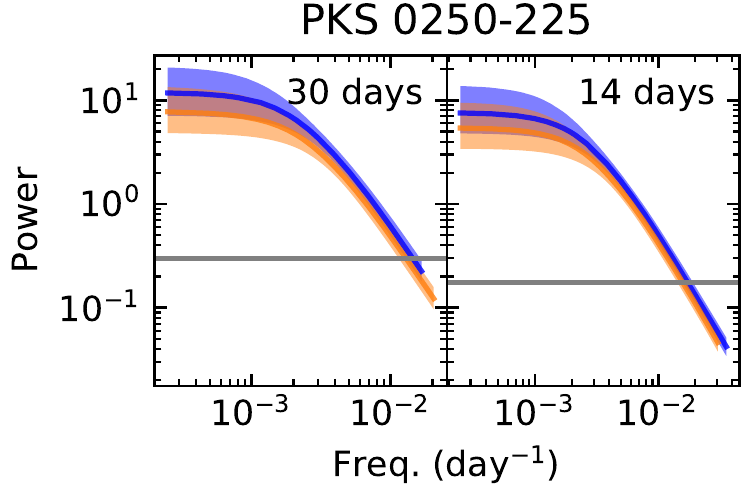}
	\includegraphics[width=0.33\textwidth]{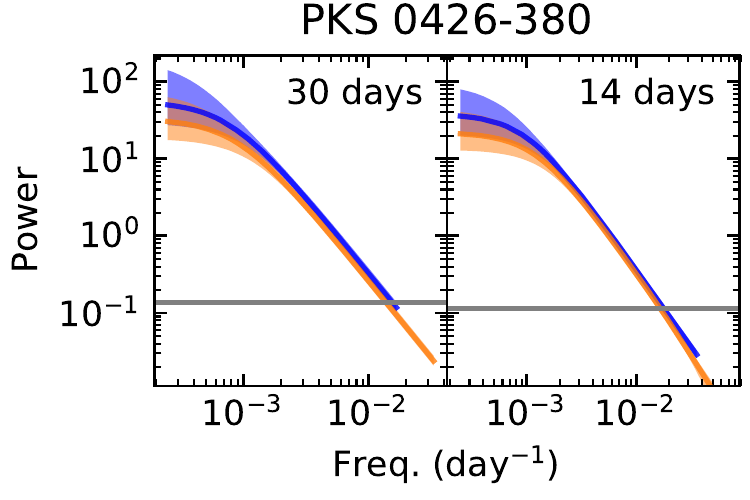}
	\includegraphics[width=0.33\textwidth]{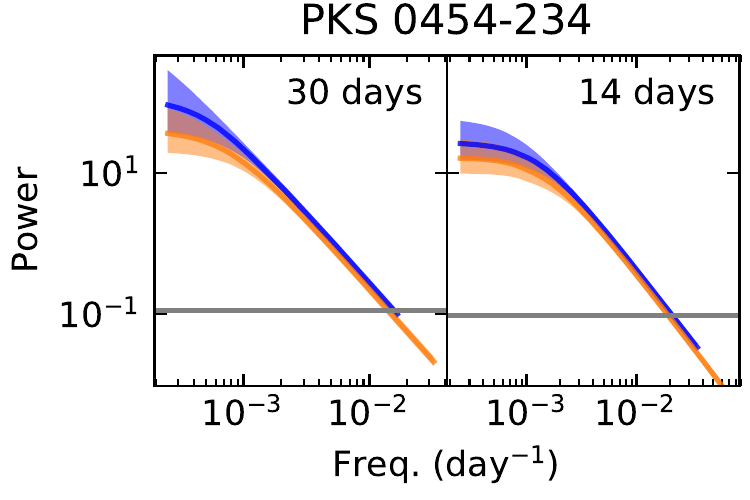}
	\includegraphics[width=0.33\textwidth]{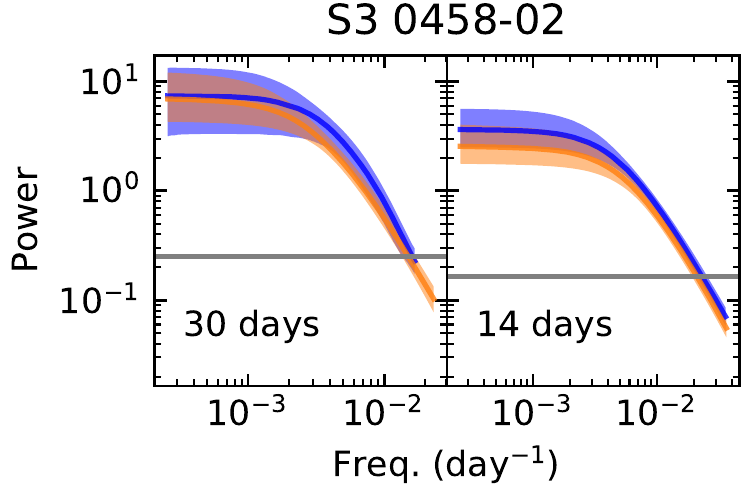}
	\includegraphics[width=0.33\textwidth]{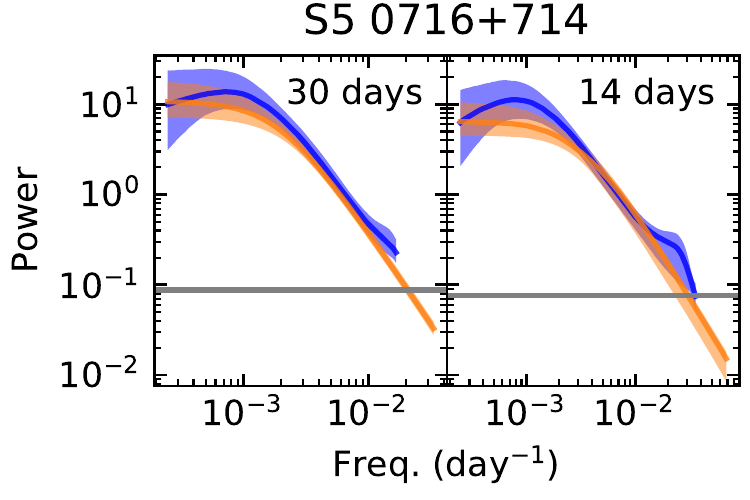}
	\includegraphics[width=0.33\textwidth]{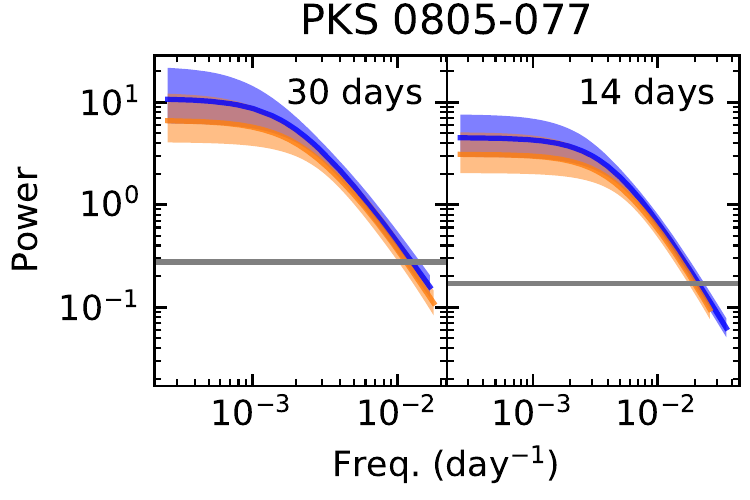}
	\caption{PSDs of the light curves of logarithmic flux for the 6 sources. The symbols and lines are the same as those in Figure \ref{fig:allpsd}.\label{appendfig:psds}}
\end{figure*}

The overfitting problem in our results (Section \ref{subsec:modeling_lc}) may be mitigated 
by reducing the wight of the flares in the whole variability. 
Here, we reanalyze the 6 sources (PKS 0250-225, PKS 0426-380, PKS 0454-234, S3 0458-02, S5 0716+714 and PKS 0805-077) using CARMA and \textit{celerite} after taking logarithm for flux. 
The model selection and fitting analysis procedures are the same as that described in Section \ref{sec:result}. 
The fitting results comparing to Figure \ref{fig:fitlc} are shown in Figure \ref{appendfig:fitresult}. One can see that all of the standardized residuals are consistent with the normal distribution. 
The PSDs of the 6 sources are also given in Figure \ref{appendfig:psds}. 
The PSDs are slightly different with those of non-logarithmic flux at high frequencies. 
This is because that modeling the light curve with logarithmic flux weakens the power of the flares and then change the PSD shape at high frequencies \citep{2019ApJ...885...12R}  .  


\end{document}